\documentclass[preprint,12pt]{elsarticle}
\biboptions{sort&compress}
\usepackage{verbatim} % comentarios
\usepackage{epsfig}
\usepackage{amssymb} 
\usepackage{mathtools}
\usepackage{lineno}
\usepackage{tikz}        %permite rellenar las figuras (circulos)
\usepackage{etoolbox}
\usepackage{ulem}         %permite el tachado de texto.
\usepackage{textcomp}

\usepackage{silence}
\usepackage{hyperref}
\usepackage{float}
\usepackage{subcaption}
\usepackage{graphicx}
\usepackage{placeins}
\begin{document}

\newcommand*{\hwplotB}{\raisebox{3pt}{\tikz{\draw[red,dashed,line 
width=3.2pt](0,0) -- 
(5mm,0);}}}

\newrobustcmd*{\mydiamond}[1]{\tikz{\filldraw[black,fill=#1] (0.05,0) -- 
(0.2cm,0.2cm) -- (0.35cm,0) -- (0.2cm,-0.2cm) -- (0.05,0);}}

\newrobustcmd*{\mytriangleright}[1]{\tikz{\filldraw[black,fill=#1] (0,0.2cm) -- 
(0.3cm,0) -- (0,-0.2cm) -- (0,0.2cm);}}

\newrobustcmd*{\mytriangleup}[1]{\tikz{\filldraw[black,fill=#1] (0,0.3cm) 
-- (0.2cm,0) -- (-0.2cm,0) -- (0,0.3cm);}}

\newrobustcmd*{\mytriangleleft}[1]{\tikz{\filldraw[black,fill=#1] (0,0.2cm) -- 
(-0.3cm,0) -- (0,-0.2cm) -- (0,0.2cm);}}
\definecolor{Blue}{cmyk}{1.,1.,0,0} 

\begin{frontmatter}

\title{Microscopic dynamics of escaping groups\\
through an exit and a corridor }

\author[add1]{E.A.~Rozan}
 \address[add1]{Departamento de F\'\i sica, Facultad de Ciencias 
Exactas y Naturales, \\ Universidad de Buenos Aires, \\
Pabell\'on I, Ciudad Universitaria, 1428 Buenos Aires, Argentina.}

 \author[add2]{G.A.~Frank}
 \address[add2]{Unidad de Investigaci\'on y Desarrollo de las 
Ingenier\'\i as, Universidad Tecnol\'ogica Nacional, Facultad Regional Buenos Aires, Av. Medrano 951, 1179 Buenos Aires, Argentina.}

 \author[add1]{F.E.~Cornes}
 \author[add1]{I.M.~Sticco}
 
\author[add1,add3]{C.O.~Dorso\corref{cor1}}%
 %\cortext[cor1]{codorso@df.uba.ar}

 \address[add3]{Instituto de F\'\i sica de Buenos Aires,\\
Pabell\'on I, Ciudad Universitaria, 1428 Buenos Aires, Argentina.}

\begin{abstract}
This research explores the dynamics of emergency evacuations in the presence of social groups. The investigation  was carried out in the context of the basic Social Force Model (SFM). We included attractive feelings between people belonging to the same social group, as proposed in Ref.~\cite{FrankDorso}. We focused on the escaping dynamics through an emergency exit and through a corridor. 
We confirmed the results appearing in Ref.~\cite{FrankDorso} for a desired velocity of $4\,$m/s, but further extended the analysis on the emergency exit to the range 1-8$\,$m/s. We noticed that the presence of groups worsens the evacuation performance. However, very strong feelings can improve the escaping time with respect to moderate feelings. We call this phenomenon “Closer-Is-Faster”, in analogy to the “Faster-Is-Slower” effect. The presence of social groups also affects the dynamic within a corridor by introducing an additional slow-down in the moving crowd.
\end{abstract}

\begin{keyword}
Evacuation \sep Social Force Model (SFM) \sep Social Groups \sep Clogging Delays

%% PACS codes here, in the form: \PACS code \sep code

\PACS 45.70.Vn \sep 89.65.Lm

%% MSC codes here, in the form: \MSC code \sep code
%% or \MSC[2008] code \sep code (2000 is the default)

\end{keyword}

\end{frontmatter}

%\linenumbers

\setlength{\parskip}{12pt}
\section{Introduction} \label{section:intro}

The basic Social Force Model (SFM) introduced by Hebling \textit{et al.} \cite{helbing} has been used to describe many emergency scenarios \cite{FrankDorso,two_exits,evacuation_obstacle,limited_visibility,cornes_2017}. The model first succeeded to explain why the crowd dynamics slows down as pedestrians try harder to escape through a narrow exit \cite{helbing,microscopic_dynamics,morphological}. This is known as the “Faster-Is-Slower” phenomenon, and it has been reported to occur in very crowded environments \cite{flow_narrow_doors,experimental_fis,parameter_optimization}

Recent research shows that the “Faster-Is-Slower” phenomenon is not the only  one occurring at high pressure levels \cite{HAGHANI2019}. As pressure undergoes dangerous levels, the crowd is supposed to behave as a compact human cluster because of the friction among the pedestrians. This is known as the “Faster-Is-Faster” regime~ \cite{microscopic_dynamics,beyond_fis,cornes2020,flow_narrow_doors, delays_nature}.

The basic SFM deals with the collective behavior of pedestrians under panic, but lacks for “\textit{social cohesion}”. That is, the behavioral effects at the “\textit{group level}” \cite{SantosAguirre}. The absence of these effects in the SFM can lead to the wrong estimation of the time required for an emergency evacuation, according to experimental research. For instance, Refs.~\cite{empirical_study,escaping_couples_facilitates} report that the egress of previously assigned groups reduces the evacuation time. However, Refs.~\cite{haghani_decision,effects_groups_gender} report that the evacuation is slightly slowed down by the presence of groups. The matter seems to be under discussion,  but it is clear that a better understanding of the consequences of social cohesion is still necessary.

Analyzes from video cameras confirm the fact that large groups have the tendency to move in spatial patterns in order to facilitate the voice communication between members~\cite{patterns,zanlungo_patterns}. These patterns occur in the context of normal walking conditions, while there is not enough evidence for this behavior under disturbing conditions (to our knowledge). This suggests, however, that group behavior fits differently into crowd models than individualistic ones. Refs.~\cite{patterns,zanlungo_patterns,virtual,review_groups} propose intra-group forces within the SFM context in order to mimic the group behavioral patterns.

Ref.~\cite{FrankDorso} studied the dynamics of dyads (groups of two members) in the context of an emergency evacuation. They introduced an attractive force between partners, additional to the usual socio-psycological forces appearing in the basic SFM (see Ref.~\cite{panic_responses} for details on this matter). The force intensity attained the emotional intimacy in the dyad. The investigation reports that the feelings of closeness play a relevant role in the evacuation time, and can yield to a significant loss in the time performance. 

Our main objective is to extend the above investigation to two quite different scenarios: escaping through a bottleneck, and passing along a straight corridor in a unidirectional flow. Both scenarios attain highly dense situations. We will limit the corridor investigation, though, to non-panicking situations for a better comparison with experimental data. More specifically, we will compute the \textit{fundamental diagram} (\textit{i.e.} the relationship between the density and the pedestrian flow), and qualitatively compare our results with empirical measurements from Ref.~\cite{helbing_FD}.

The investigation is organized as follows. In Section~\ref{section:background} we present the highlights of the SFM and its extension to the group model, as defined in Ref.~\cite{FrankDorso}. We also include definitions for clustered structures and the fundamental diagram. In Section~\ref{section:experimental} we compare the extended SFM with the (scarce) empirical data on social groups. Section \ref{section:numsim} details the simulation procedures for studying the bottleneck and the corridor scenarios. In Sections~\ref{section:t_evac}, \ref{section:delays} and \ref{section:intragroup} we present the results corresponding to the bottleneck scenario, while Section \ref{section:results_fd} shows the corresponding results for the corridor. A discussion on the results is shown in Section~\ref{section:discussion}. Our main conclusions can be found in Section~\ref{section:conclusions}.

\section{Theoretical Background} \label{section:background}
\subsection{Social Force Model (SFM)}\label{section:SFM}

Our research was carried out in the context of the Social Force Model (SFM) proposed by Helbing \textit{et al.}~\cite{helbing}. This model states that human motion is motivated by the desire of people to reach a certain destination, and can be further affected by environmental factors. The basic SFM considers the following equation of motion for any pedestrian $i$:

\begin{equation}\label{SFM}
m_i\,\displaystyle\frac{d\mathbf{v}_i}{dt}(t) = \mathbf{f}_d^{(i)}(t) +
\displaystyle\sum_{j}\displaystyle\mathbf{f}_s^{(ij)}(t)+\displaystyle\sum_{j} \mathbf{f}_g^{(ij)}(t)
\end{equation}

\noindent where $j$ represents any other pedestrian or a wall.

Three kind of forces are included in the above equation: the desire force, the social force and the sliding friction force. An additional force will be introduced in Section~\ref{Fa_section}.

The “desire force” $\mathbf{f}_d$ corresponds to the pedestrians own willings to move at the desired velocity $v_d$. %But, in order to reach the desired target, they need to accelerate (decelerate) from their current velocity $\mathbf{v}(i)(t)$. This acceleration (or deceleration) represents a “desire force” since it is motivated by their own willingness. 
It is associated with the required acceleration (or deceleration) from their current velocity $\mathbf{v}_i(t)$ to the desired one. The corresponding expression is as follows

\begin{equation}
\mathbf{f}_d^ {(i)}(t)=m_i\,\displaystyle\frac{v_d^{(i)}(t)\  \hat{\mathbf{e}}_d^ {(i)}(t)-\mathbf{v}_i(t)}{\tau}\label{Fd}
\end{equation}

\noindent where $m_i$ is the mass of the pedestrian $i$ and $\tau$ represents the relaxation time needed to reach his (her) desired velocity. $\hat{\mathbf{e}}_d$ is the unit vector pointing to the target position, while $v_d$ attains the anxiety level of the pedestrian. For simplicity, we will assume that $v_d$ remains unchanged during the entire process and is the same for all of the individuals, but $\hat{\mathbf{e}}_d$ changes according to the current position of the pedestrian. 

In the context of an evacuation process, if no acquittance, friendship or family engagements exist, the most common tendency is to keep some space between each other, or from the walls. The “social force” $\mathbf{f}_s^{(ij)}$ represents this socio-psychological tendency between any two pedestrians, say $i$ and $j$, in order to preserve their \textit{private space}. It is assumed to be

\begin{equation}
\mathbf{f}_s^{(ij)}=A_i\,e^{(r_{ij}-d_{ij})/B_i}\mathbf{n}_{ij}\label{Fs}
\end{equation}

\noindent where $A_i$ and $B_i$ are two fixed parameters. $r_{ij}$ = $r_i+r_j$ is the sum of the pedestrians radius (say, the shoulder-neck distance), and $d_{ij}$ is the distance between their centers of mass. $\mathbf{n}_{ij}$ stands for the unit vector in the $\vec{ji}$ direction. If $j$ represents a wall, then $d_{ij}$ corresponds to the shortest distance between the pedestrian and the wall, and $r_j$ is set to zero.

The emotional reactions due to friendship or family engagements may also be handled as a “socio-psychological force”. We will discuss this matter in Section~\ref{Fa_section} below.

The sliding friction is present whenever two individuals (or an individual and the wall) get in contact. It is known as the “granular force” within the SFM, and is considered to be a linear function of the relative tangential velocities of the contacting individuals. %In the case of the friction exerted by the wall, the force is a linear function of the pedestrian tangential velocity, as $\mathbf{v}_j=0$ for the wall. 
Its mathematical expression is as follows

\begin{equation}
\mathbf{f}_g^{(ij)}=\kappa\,g(r_{ij}-d_{ij})\,\left( \Delta \mathbf{v}_{ij}\cdot \: \hat{\mathbf{t}}_{ij}\right) \: \hat{\mathbf{t}}_{ij} \label{Fg}
\end{equation}

\noindent where $\kappa$ is the friction coefficient. The function $g(r_{ij}-d_{ij})$ is equal to its argument when this is positive (that is, when $d_{ij}<r_{ij}$, meaning that pedestrian $i$ and $j$ are closer than the contact distance) and equals zero for any other case. $\Delta \mathbf{v}_{ij}\cdot  \:\hat{\mathbf{t}}_{ij}$ represents the difference between the tangential velocities of the sliding bodies (if $j$ represents a wall, then $\Delta \mathbf{v}_{ij}=\mathbf{v}_i$). The whole set of parameters' values can be found in Section~
\ref{section:numsim_general}. 
\subsection{The Attractive Force} \label{Fa_section}

Ref.~\cite{FrankDorso} introduced an attractive force into the equation of motion in order to mimic the behavioral patterns of groups. For instance, when one of the members of the group is pushed aside, he (she) will try to move back to the the space in common within the group. This behavior is considered in Ref.~\cite{FrankDorso} to be associated with attractive feelings among the members of the group, and are supposed to be balanced with the “private sphere” preservation. 

The attractive force in Ref.~\cite{FrankDorso} is assumed to derive from a Fermi-like potential. Its mathematical expression reads as follows

\begin{equation}
    \mathbf{f}_a^{(ij)}= -\frac{\epsilon}{4D_i}\cosh ^{-2}
    \left( \frac{C_i-d_{ij} }{2D_i} \right)  \mathbf{n}_{ij}\label{Fa}
\end{equation}

\noindent where $\epsilon$ represents the intensity of the attraction and will vary depending on the the “closeness” to the group members. $C_i$ and $D_i$ are fixed parameters, set to $D_i=0.5B_i$ and $C_i=r_{ij}+7B_i$ in Ref.~\cite{FrankDorso}. We will hold these values throughout our research. 

\begin{figure}[ht]\centering
 \includegraphics[width=0.6\linewidth]{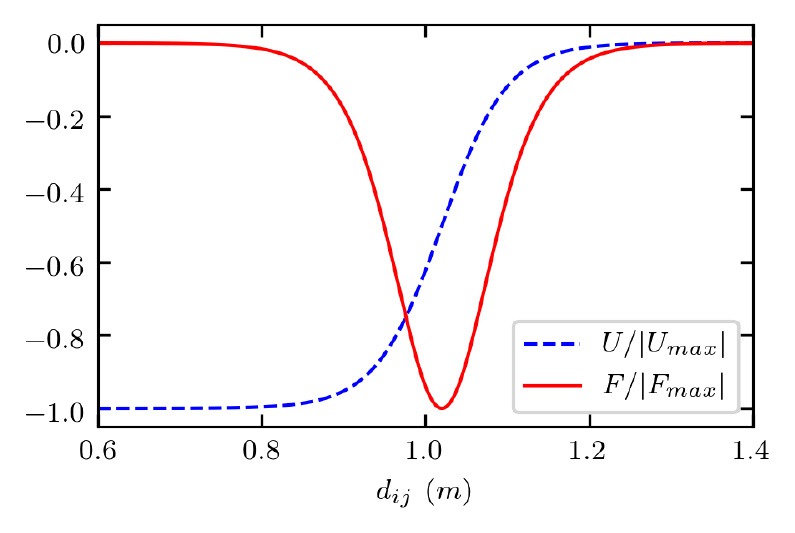}
 \caption{Attractive potential ($U$) and force ($F=-\frac{\mathrm{dU}}{\mathrm{d}r}$). The parameters values are $C$~=~$1.02$~m and $D=0.04$~m. Note that the curves are normalized and therefore do not depend on the value of $\epsilon$.}
 \label{fig_atrac}
\end{figure}

Fig.~\ref{fig_atrac} shows the attractive potential and the force for the set of parameters used in the present investigation (see caption for details).

The feeling intensity can be as strong as $\epsilon=10^9\,$N$\cdot$m, as reported in Ref.~\cite{FrankDorso}. Thus, we will express our results in the more convenient variable defined as follows:
\begin{equation}
\varepsilon = \log_{10}\left(\frac{\epsilon}{\text N\cdot\text m}\right)
\end{equation}

\subsection{Clustering structures}

Human clusters arise when pedestrians get in contact between each other and are responsible for the time delays during the evacuation process ~\cite{microscopic_dynamics,morphological,beyond_fis,cornes2020,two_exits}. Clusters of pedestrians can be defined as the set of individuals that for any member of the group (say, $i$) there exists at least another member belonging to the same group ($j$) in contact with the former. Thus, we define a spatial cluster following the mathematical formula~\cite{fragment_recognition}

\begin{equation}\label{granular_cluster}
   C_g: P_i \in C_g  \Leftrightarrow \exists\ P_j \in C_g \:/\:d_{ij}<r_i+r_j
\end{equation}

\noindent where $p_i$ represents the $i$th pedestrian, $r_i$ their radius (shoulder-neck). %This means that $C_g$ is a set of pedestrian that interact via the friction force (Eq.~(\ref{Fg})) in addition to the social and attraction forces (Eqs.~(\ref{Fs}) and ~(\ref{Fa}), respectively). 

\begin{figure}[ht]\centering
 \includegraphics[width=0.5\linewidth]{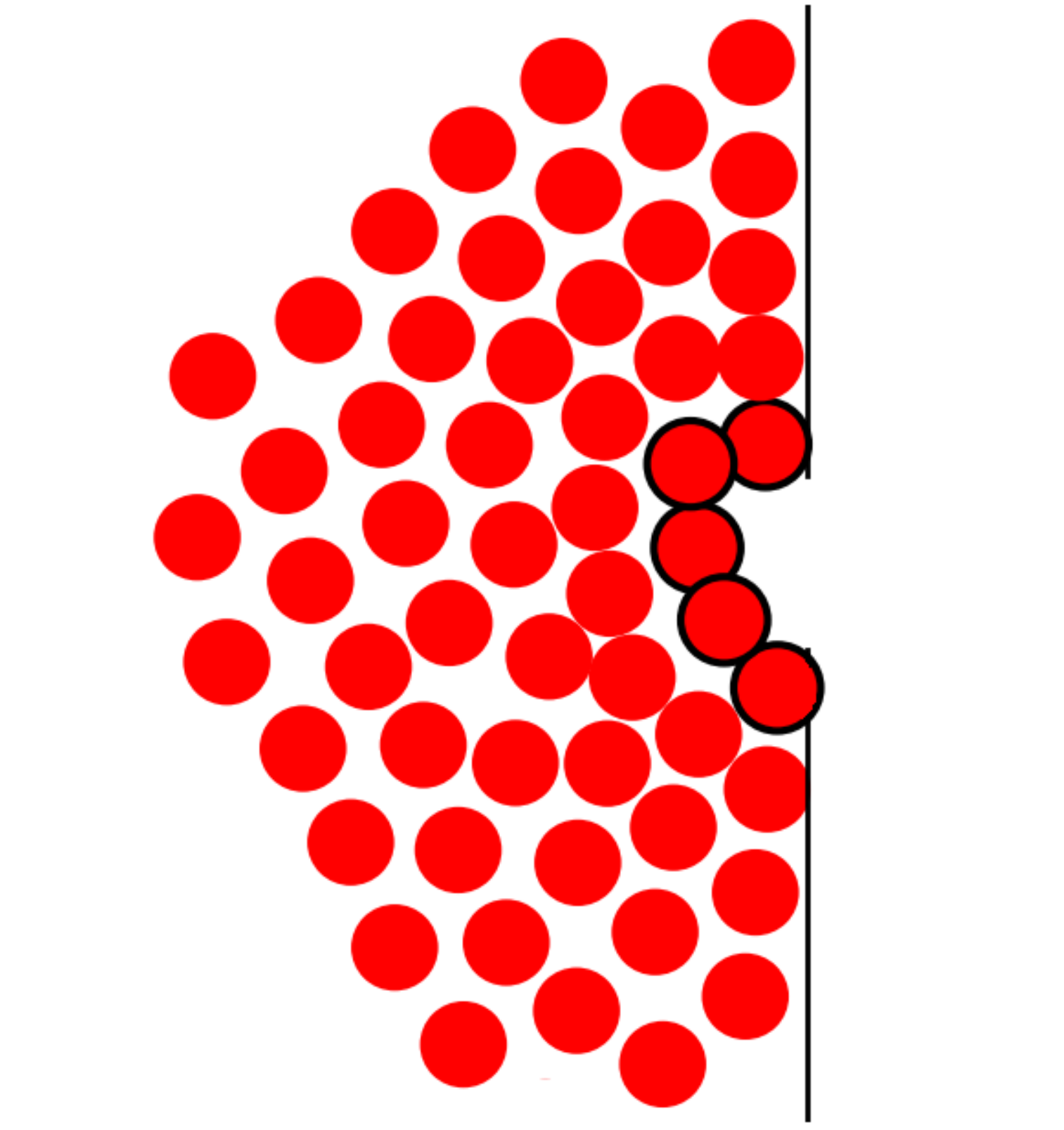}
  \caption{Example of a blocking cluster formed during an evacuation process. Individuals are represented as red circles, and those with black borders are in contact, thus they are part of a cluster. As one of them is in contact with the upper wall and another one is in contact with the bottom one, this set of pedestrians is a blocking cluster. }  
\label{cluster_example}
\end{figure}

From all granular clusters, the \textit{blocking clusters} are the minimum set of contacting pedestrians (belonging to a cluster) that connects the walls on both sides of the exit. Roughly speaking, it refers to the shortest chain of contacting pedestrians that links both sides of the exit door. An example of a blocking cluster is shown in Fig.~\ref{cluster_example}.

\subsection{Fundamental Diagram}\label{section:fd_intro}
Many researches on pedestrian dynamics focus their attention on the relation between the flow and the density of a moving crowd. This relation is represented by the “fundamental diagram” and it has become one of the most common ways to characterize the pedestrians’ dynamics along a corridor~\cite{role_friction,helbing_FD,unidirectional_dense_crowd,seyfried}. The flow is defined as

\begin{equation}
\mathbf{J}=\rho \mathbf{V}
\label{flow_def}
\end{equation}

\noindent where $\rho$ represents the local density (say, the number of pedestrians within a region divided by its area) and $\mathbf{V}$ corresponds to the local velocity (the averaged velocity of the pedestrians in that region).

Details on empirical and simulated fundamental diagrams and measurement conditions can be found in Refs.~\cite{helbing_FD,role_friction}.

\section{Experimental background}\label{section:experimental}
In this section, we fit our numerical simulations into the available experimental data, in order to obtain estimates of $\varepsilon$ (the intensity of the attractive force) for each category of social group. 

\begin{figure}[ht]\centering
 \includegraphics[width=0.8\linewidth]{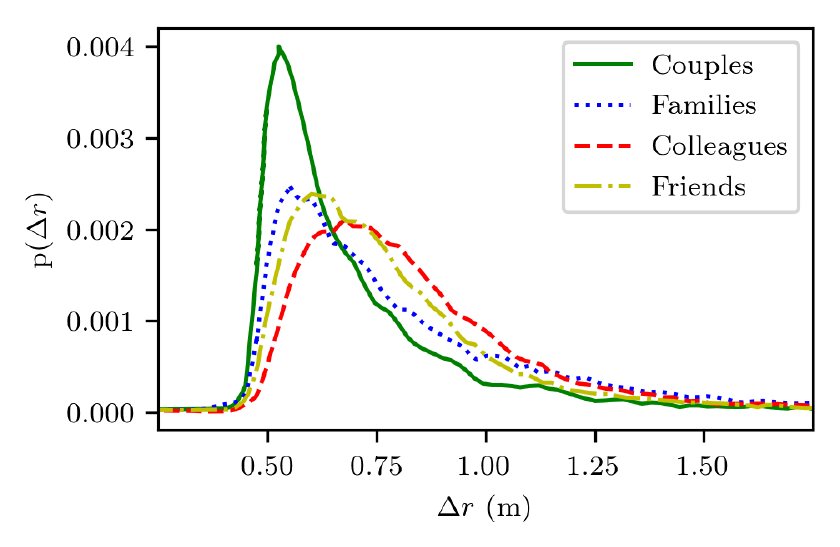}
 \caption{(Color online only) Probability distribution function for the distance ${\Delta r=|\mathbf{r}_1-\mathbf{r}_2|}$ within a group. Each curve corresponds to a different group category. These recordings were taken in “The Asia and Pacific Trade Center” in Osaka (see Ref.~\cite{dependence_dyad_dynamics} for details). }
 \label{dyad_distribution}
\end{figure}

Ref. \cite{dependence_dyad_dynamics} captured, by means of video cameras, the walking characteristics of group members. The authors classified this information into group categories, and determined the distance distribution function between members of dyads (that is, groups of two individuals). Their results can be seen in Fig.~\ref{dyad_distribution}. 

%\FloatBarrier

We are now able to link the expected distances for \textit{couples} or \textit{colleagues} with the intensity of the attraction feelings. We proceed as follows: first, we compute the equilibrium distances for a wide range of intensity values $\varepsilon$ (see \ref{appendix:balance} for details). Secondly, we compare these distances with the corresponding values for \textit{couples} or \textit{colleagues}. Finally, we associate an intensity range for $\varepsilon$ within each category.

While all the experimental distributions in Fig.~\ref{dyad_distribution} are quite similar, the less involving categories display wider distributions. For instance, the most probable distances  for the \textit{couples} category ranges from 0.5 m to 0.65 m ($\pm$ one standard deviation). The most probable distances for the \textit{colleagues}, instead, ranges from 0.6 m to 0.9 m ($\pm$ one standard deviation). This kind of dispersion makes it difficult to classify groups when the walking distance is around $\Delta r\sim 0.6\,$m.  Thus, we will not try to fit the data for the intermediate categories (\textit{family} and \textit{friends}), in order to avoid ambiguities.

\begin{figure}[ht]\centering
 \includegraphics[width=0.8\linewidth]{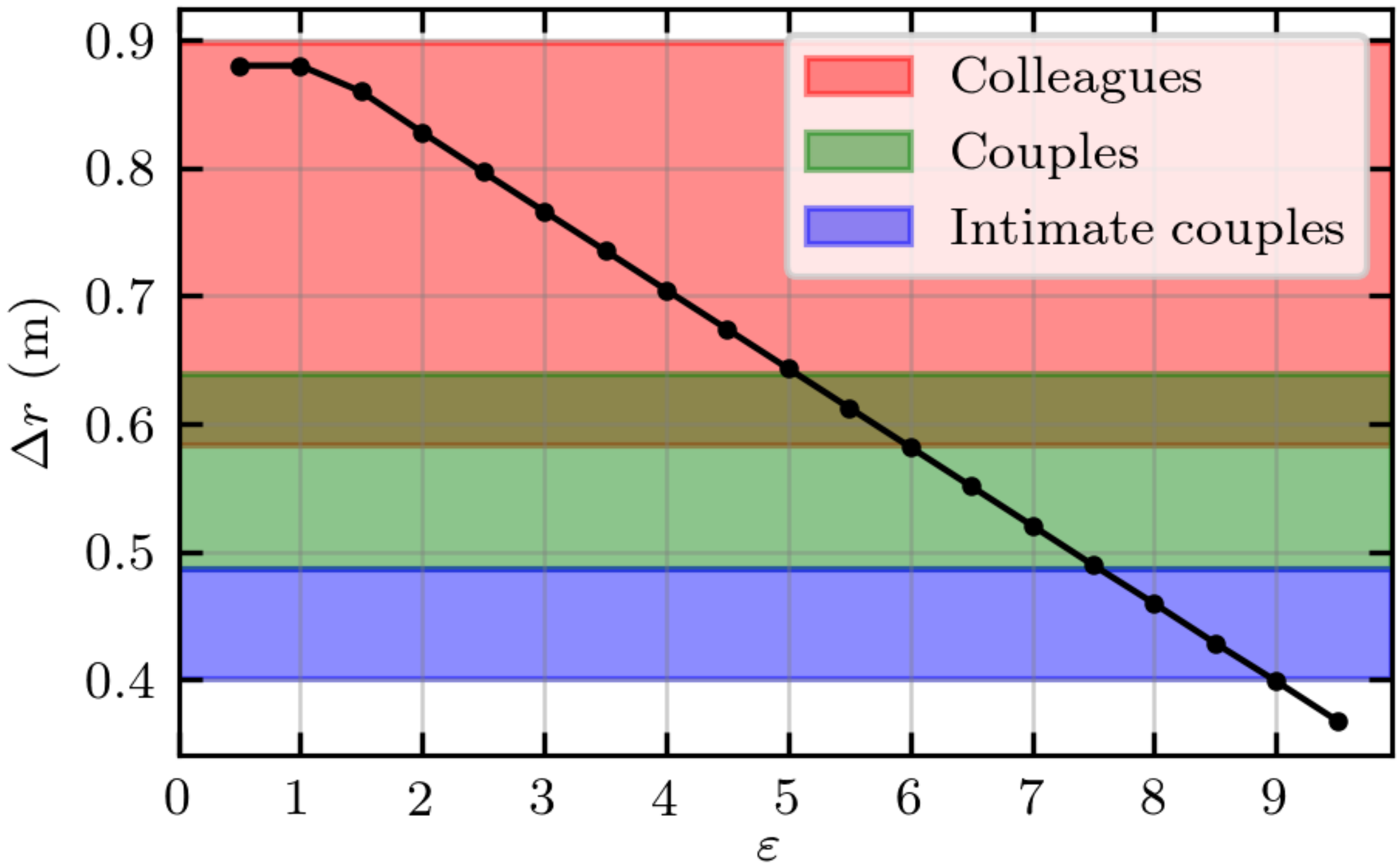}
 \caption{(Color online only) Equilibrium distance between both members of a walking dyad in a low density corridor, as a function of the attractive intensity $\varepsilon$. Details on the parameters used for the SFM can be found in Section~\ref{section:numsim} below. The red and green shaded regions correspond to the most probable distances for \textit{colleagues} and \textit{couples}, respectively, while the blue shaded region corresponds to the presumed \textit{intimate couples} category (see text for details).}
 \label{exp_results}
\end{figure} 

Fig.~\ref{exp_results} shows the equilibrium distance of a walking dyad as a function of the attractive force intensity. Notice that a linear relation exists between $\Delta r$ and $\varepsilon$. The shaded regions in Fig.~\ref{exp_results} indicate the mapping from $\varepsilon$ to  the \textit{couple} or \textit{colleagues} category (see caption for details). We see that the range of $\varepsilon$ corresponding  to \textit{colleagues} is $1<\varepsilon<6$, and the one corresponding to \textit{couples} is $5<\varepsilon<7.5$.

Notice that if the attraction feelings $\varepsilon$ lie between 5 and 6, we are not able to tell precisely if they are \textit{colleagues} or a \textit{couple}. This requires additional information that is out of the scope of our investigation.

Besides, note from Fig.~\ref{dyad_distribution} that all the distribution functions vanish for distances shorter than 0.5$\,$m. We presume that this is a consequence of cultural habits (say, japanese culture in the case of Fig.~\ref{dyad_distribution}). We leave open the possibility that other cultures might allow distances shorter than 0.5$\,$m (hugging, holding hands, etc.). We will refer to dyads attaining distance intervals between 0.4 and 0.5$\,$m as \textit{intimate couples}. This distance range maps to $\varepsilon$ between 7.5 and 9 in Fig~\ref{exp_results}.

\section{Numerical Simulations} \label{section:numsim}
We first present the general simulation conditions. We then go over the specific conditions for the bottleneck and the straight corridor situations separately. 

\subsection{General Simulation Conditions}\label{section:numsim_general}

Pedestrians were modeled as soft spheres. Initially, the individuals were randomly distributed along the simulation box with random initial velocities, following a Gaussian distribution with null mean value. At the beginning of the simulation, group members were separated by a random distance between 0.4 and 0.7$\,$m and placed in a random orientation. The desired velocity $v_d$ was the same for all the individuals, and the attractive feelings intensity $\varepsilon$ was the same for all of the groups in each simulation process. 

We set the SFM parameters to the usual literature values, as shown in Table~\ref{parameters} (see details in Refs.~\cite{FrankDorso,helbing,two_exits,handbook}). Notice that two friction coefficients appear, which correspond to the commonly accepted values in the literature. However, researchers did not arrive to a unique value for the friction at bottlenecks or corridors~\cite{role_friction,parameter_optimization}. We will accept two possible values, without introducing a discussion on this matter here.  

\begin{table}[ht]
\centering
\begin{tabular}{|l|l|l|}
\hline
\textbf{Parameter} & \textbf{Value}                       &     \textbf{Meaning}         \\ \hline
$m$                & 70 Kg                                & Pedestrian's mass \\
$r$                & 0.23 m                               & Pedestrians' radius (shoulder-neck) \\
$\tau$             & 0.5 s                                & Relaxation time \\
$A$                & 2000 N                               & Intensity of $\mathbf f_s$ \\
$B$                & 0.08 m                               & Characteristic length of $\mathbf f_s$ \\
$C$                & 1.02 m                               & Maximum of $\mathbf f_a$ \\
$D$                & 0.04 m                               & Halved characteristic length of $\mathbf f_a$ \\
$\kappa$           & $2.4\cdot 10^5\,$Kg$\,$m$^{-1}\,$s$^{-1}$ & Friction coefficient in bottlenecks \\
$\kappa_\text{augmented}$           & $1.2\cdot 10^6\,$Kg$\,$m$^{-1}\,$s$^{-1}$ & Friction coefficient in corridors\\ \hline
\end{tabular}
\caption{Parameters used for the SFM simulations. Notice that there are two different values for $\kappa$ (see text for details).}
\label{parameters}
\end{table}

Each simulation process had two kinds of individuals: single pedestrians and group members. Single individuals interact with others through the social and friction forces ($\mathbf f_s$ and $\mathbf f_g$, respectively). Group members, additionally, are mutually attracted by the attractive force $\mathbf f_a$ (see Section~\ref{section:SFM} and \ref{Fa_section} for details).

The simulations were performed using \textsc{Lammps} molecular dynamics simulator with parallel computing capabilities~\cite{lammps}. The time integration algorithm  followed the velocity Verlet scheme with a time step of $10^{-4}\,$s. We implemented special modules in C++ for upgrading the \textsc{Lammps} capabilities to perform the SFM simulations. Data recordings were done at time intervals of $0.05\,$s, that is, at intervals as short as 10\% of the pedestrian's relaxation time $\tau$. 

\subsection{Evacuation in the bottleneck scenario}

\begin{figure}[ht]\centering
 \includegraphics[width=\linewidth]{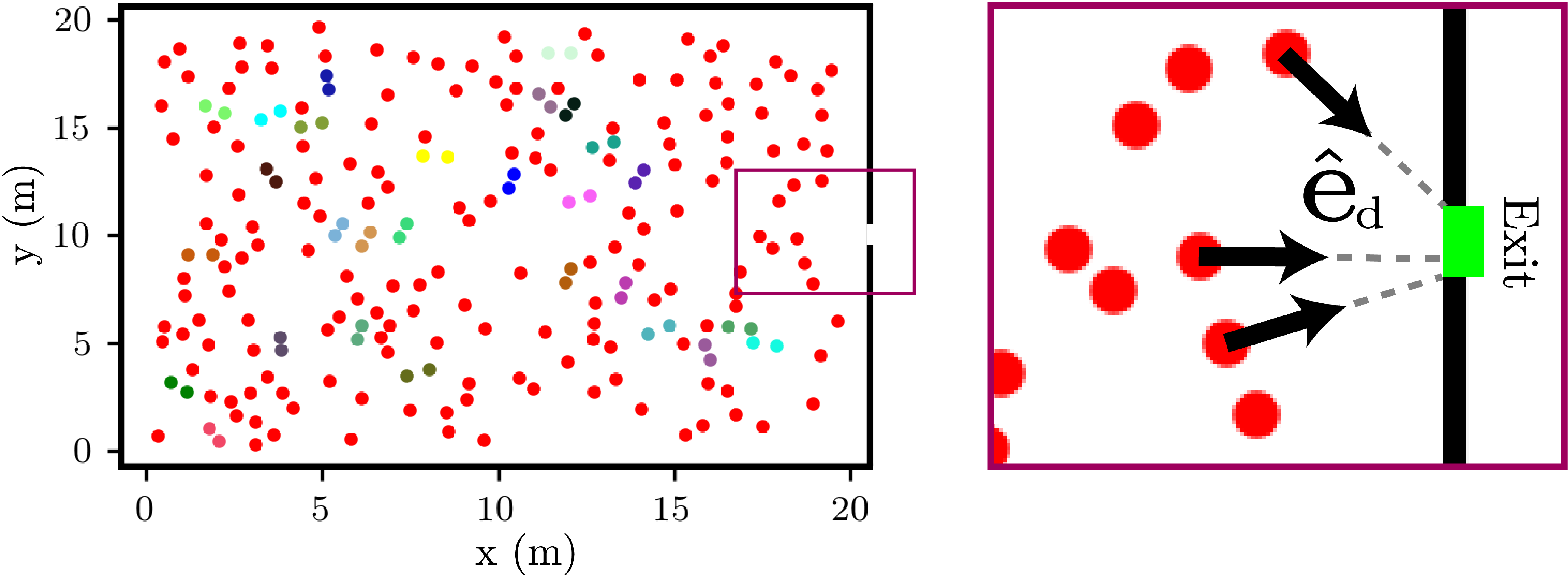}
 \caption{(Color online only) Initial configuration for an evacuation process through a bottleneck. 225 pedestrians are located in a 20$\,$m $\times$~20$\,$m room with a single door. Single individuals are represented as red circles, while each dyad is represented by a other colors (online version only). $\mathbf{e}_d^{(i)}$ points directly to the exit.}
 \label{IC_example}
\end{figure}

We simulated evacuation processes from a 20$\,$m $\times$~20$\,$m room with 225 pedestrians. The room had a single exit placed symmetrically, as shown in Fig.~\ref{IC_example}. The door width was 0.92$\,$m (equal to the diameter of two individuals), as in Refs.~\cite{FrankDorso,morphological,microscopic_dynamics,beyond_fis}. All the individuals had the willing to go to the exit door. That is, at each time-step the desired direction $\hat{\mathbf{e}}_d$ was updated in order to point to the exit, as depicted in Fig.~\ref{IC_example}. The simulation process lasted until 160 individuals ($\sim$70\%) left the room. If this condition could not be fulfilled in the first 3000$\,$s, the process was stopped. We ran 100 processes for each set of $v_d$ and $\varepsilon$ values, in order to get enough data for statistical analysis. 

\FloatBarrier

We focused on two cases: 25\% of the pedestrians belong to dyads (say, 56 individuals grouped in 28 dyads), or 100\% of the pedestrians  belong to dyads (224 individuals grouped in 112 dyads). The latter is supposed to be an extreme situation, but is intended as a bounding case for our results.

We did not include bigger groups in order to avoid the effects of group walking patterns.

\subsection{Straight corridor scenario}
\begin{figure}[ht]\centering
 \includegraphics[width=0.6\linewidth]{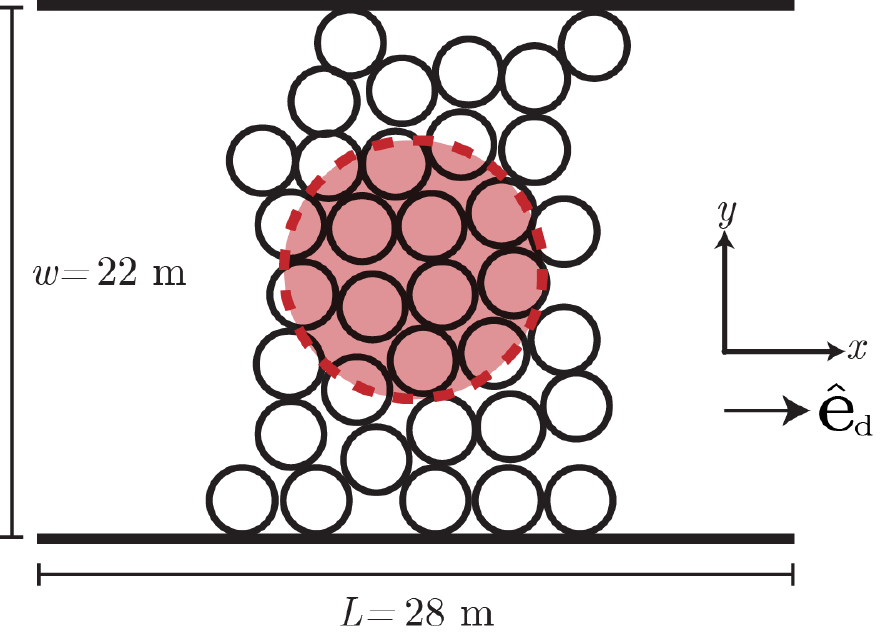}
 \caption{Schematic diagram for individuals in the corridor. White  circles represent pedestrians moving towards the positive values of $x$. The red circle in the middle corresponds to the sampling region, and its radius was set to 1$\,$m.}
 \label{fd_region}
\end{figure}

This scenario explored the flow of pedestrians along a straight corridor of width $w~=22$~m (say, similar to the width of the entrance at the Jamaraat Bridge; see Ref.~\cite{helbing_FD} for details) and length $L=28\,$m with periodic boundary conditions (see Fig~\ref{fd_region}). The global density ranged from $\rho=0.5$ to 9$\,$P/m$^2$. 

The desired velocity for every pedestrian was set to $v_d=1\,$m/s, which correspond to low anxiety levels. The desired direction $\hat{\mathbf{e}}_d$ always pointed from left to right (see Fig.~\ref{fd_region}). %This allowed the individuals to move in a unidirectional flow. 

The sampling area was located in the middle of the corridor as shown in Fig.~\ref{fd_region}. The velocity $\mathbf V$ of the individuals inside the circle was recorded every 0.5$\,$s for 80$\,$s, after the first 20$\,$s of the simulation. The flow $\mathbf J$ was further computed according to Eq.~(\ref{flow_def}).

Recall from Table~\ref{parameters} that the accepted  friction coefficient for the corridor scenario is five times higher than the one for the bottleneck. This value yields to a fundamental diagram that agrees with the experimental data appearing in Ref.~\cite{helbing_FD} (see Ref.\cite{role_friction}). Therefore, we will consider this augmented value for simulations in the corridor scenario without entering into further discussions.

In all cases, 70\% of the individuals belonged to groups. We studied situations of groups of up to 5 members in Section~\ref{section:results_fd}. However, we will not analyze the groups pattern formation or the decision-making processes.

\section{Results}\label{section:results}

\subsection{Evacuation time with social groups}\label{section:t_evac}

\begin{figure}[ht]\centering
 \begin{subfigure}{.49\textwidth} \centering
 \includegraphics[width=\linewidth]{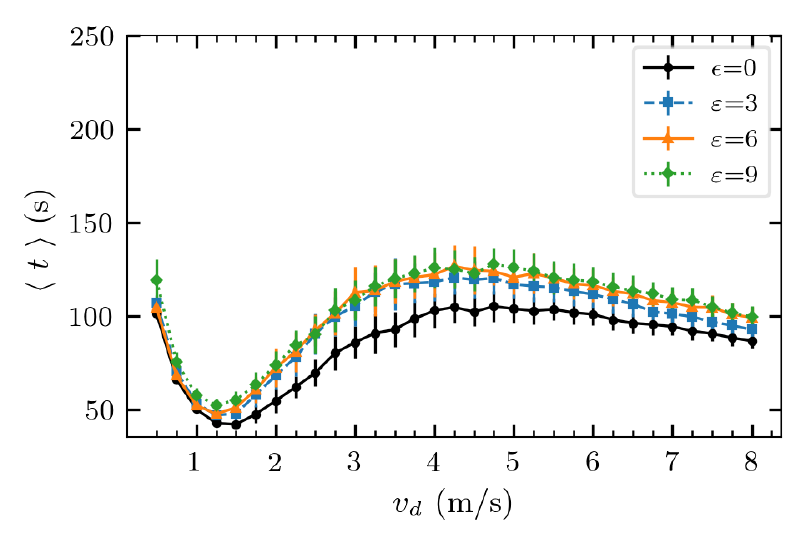}
 \caption{25\%}
  \label{t_fixed_epsilon_a}
 \end{subfigure}
 \begin{subfigure}{.49\textwidth} \centering
 \includegraphics[width=\linewidth]{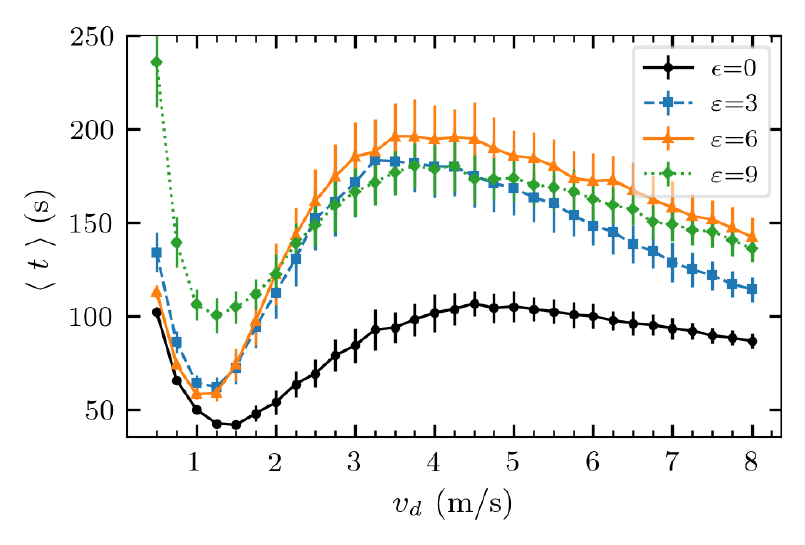}
 \caption{100\%}
 \label{t_fixed_epsilon_b}
 \end{subfigure}
\caption{(Color online) Mean evacuation time for the egress of 160 individuals as a function of $v_d$ and the attractive feelings intensity level $\varepsilon$. The door width was 0.92$\,$m (equal to two pedestrians' diameter). The black curve, $\epsilon=0$, corresponds to no dyads within the crowd. (a) 25\% of the pedestrians were grouped in dyads. (b) 100\% of the pedestrians were grouped in dyads.}
\label{t_fixed_epsilon}
\end{figure}

In this section we present the results corresponding to the bottleneck scenario. Fig.~\ref{t_fixed_epsilon} shows the mean evacuation time $\langle t \rangle$ of a crowd for a wide range of anxiety levels and attractive intensities (see caption for details). 

Fig.~\ref{t_fixed_epsilon} exhibits the usual “Faster-Is-Slower” and “Faster-Is-Faster” behaviors, as explained in the literature (see Refs.~\cite{morphological,cornes2020,beyond_fis}). However, the presence of dyads worsens the evacuation performance. This phenomenon becomes more significant as the fraction of dyads increases (see Fig.~\ref{t_fixed_epsilon_b}). 

\begin{figure}[ht]\centering
 \begin{subfigure}{.5\textwidth} \centering
 \includegraphics[width=\linewidth]{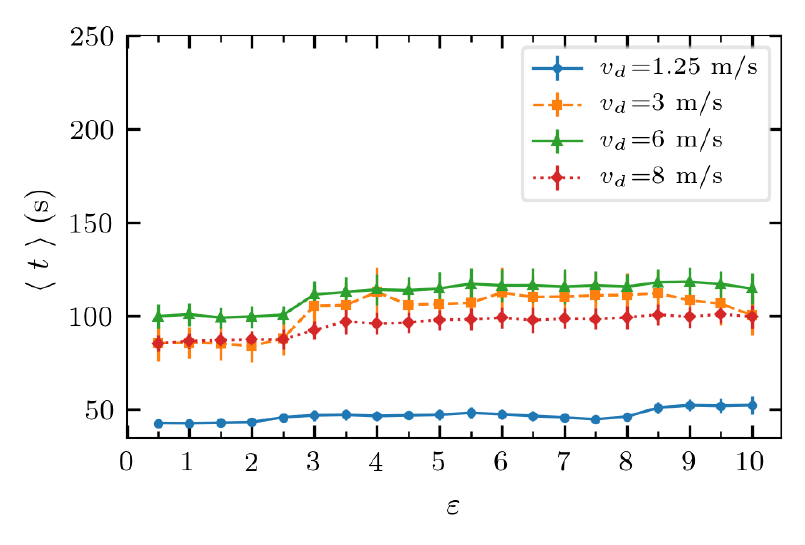}
 \caption{25\%}
  \label{t_fixed_vd_a}
 \end{subfigure}%
 \begin{subfigure}{.5\textwidth} \centering
 \includegraphics[width=\linewidth]{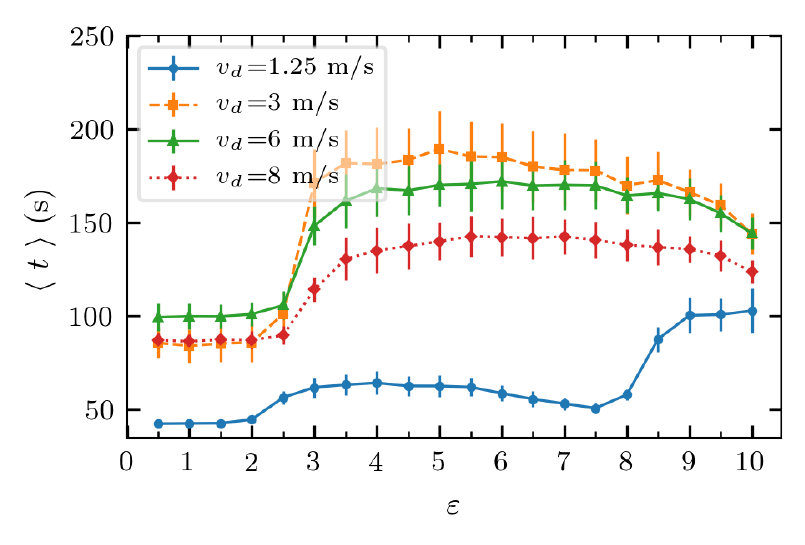}
 \caption{100\%}
 \label{t_fixed_vd_b}
 \end{subfigure}
\caption{(Color online) Mean evacuation time for 160 individuals as a function of $\varepsilon$ for fixed values of $v_d$. (a) 25\% of the pedestrians were grouped in dyads. (b) 100\% of the pedestrians were grouped in dyads. The simulation conditions were the same as in Fig.~\ref{t_fixed_epsilon}}
\label{t_fixed_vd}
\end{figure}

Our concern is with the intensity of the feelings within the dyads, $\varepsilon$. Thus, we explored the evacuation performance for four representative values of $v_d$. Fig.~\ref{t_fixed_vd} shows the corresponding mean egress time as a function of $\varepsilon$.

The very first inspection of the plots shows that the slopes are almost flat, except at intensities $\varepsilon$ between 2 and 3, and above 8. Fig.~\ref{t_fixed_vd_b} exhibits more noticeable changes at these intensities than in Fig.~\ref{t_fixed_vd_a}.

The intensities $\varepsilon$ between 2 and 3 correspond to feeling forces in the range of 625 to 6250$\,$N.  Recall that the social force approaches 2000$\,$N near the contact distance. Thus, the attractive feelings appear quite weak with respect to the social repulsion below $\varepsilon=2$, but quite strong beyond $\varepsilon=3$. The range $(2<\varepsilon<3)$ produces a significant change in $\langle t\rangle$, according to Fig.~\ref{t_fixed_vd_b}. Notice, however, that the egress time settles to a well established level above $\varepsilon=3$, depending on the degree of anxiety (say, the desired velocity $v_d$). 

The egress time performs somewhat better beyond $\varepsilon=7$ if all the individuals are grouped in dyads and $v_d>1.25\,$m/s (see Fig.~\ref{t_fixed_vd_b}). This corresponds to \textit{intimate couples} evacuating in a crowded environment. Ref.\citep{FrankDorso} reports this phenomenon for $v_d=4\,$m/s only. The authors explain that dyads are so tightly linked that the whole dyad mimics a single big person, improving the evacuation time. 

The time improvement observed for \textit{intimate couples} ($\varepsilon>7$) does not apply to the low anxiety curve in Fig.~\ref{t_fixed_vd_b} (say, $v_d=1.25\,$m/s). The reason for this is explained in \ref{appendix:lowvdhigheps}. We will not make further comments on this phenomena since it belongs to a non-panic scenario. 

The behavior of $\langle t \rangle$ as a function of $\varepsilon$ is qualitatively similar to its counterpart as a function of $v_d$ (see Fig.~\ref{t_fixed_epsilon}). We will call “Closer-Is-Slower” (CIS) and “Closer-Is-Faster” (CIF) the regimes in Fig.~\ref{t_fixed_vd} that mimic the “Faster-Is-Slower” and “Faster-Is-Faster” regimes in Fig.~\ref{t_fixed_epsilon}, respectively. That is, CIS  stands for the interval attaining a positive slope Fig.~\ref{t_fixed_vd_b} (say, $2<\varepsilon <5$) and CIF corresponds to the interval attaining a negative one ($\varepsilon >5$). Recall that the former is associated with the expected feelings among \textit{colleagues} while the latter is associated with \textit{intimate couples}.

\begin{figure}[ht]\centering
 \begin{subfigure}{.49\textwidth} \centering
 \includegraphics[width=\linewidth]{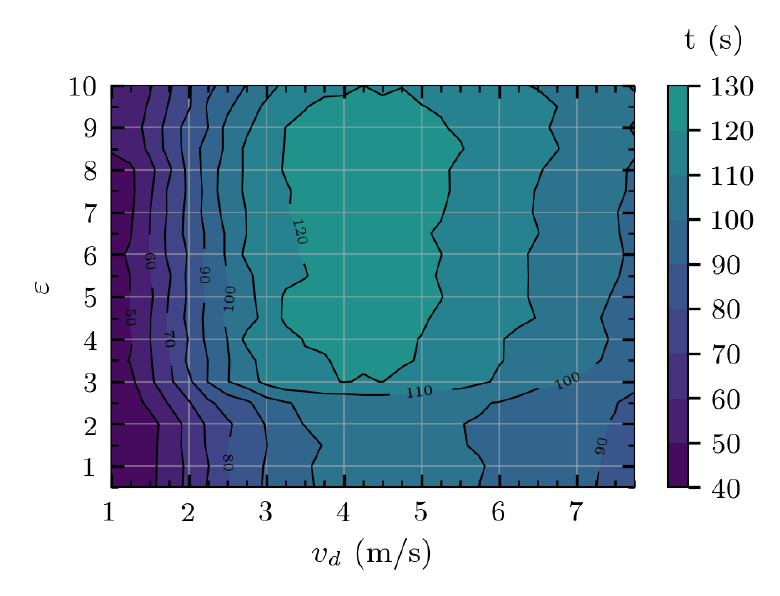}
 \caption{25\% dyads}
 \label{t_25}
 \end{subfigure}%
 \begin{subfigure}{.49\textwidth} \centering
 \includegraphics[width=\linewidth]{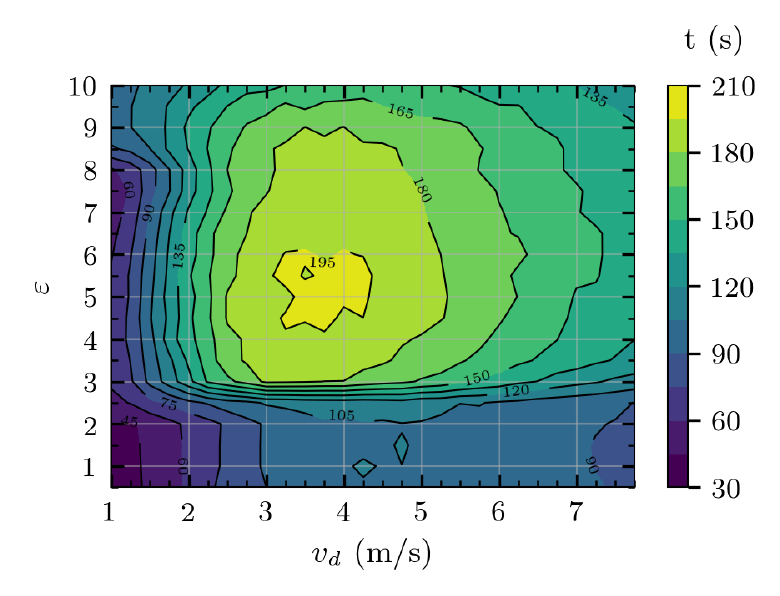}
 \caption{100\% dyads}
  \label{t_100}
 \end{subfigure}
\caption{(Color online only) Mean egress time for 160 individuals as a function of both the attractive feeling intensity $\varepsilon$ and the desired velocity $v_d$, averaged over 100 simulations. (a) 25\% of the pedestrians were grouped in dyads. (b) 100\% of the pedestrians were grouped in dyads.}
\label{t_heat_map}
\end{figure}

The contour-map for the evacuation time as a function of both $v_d$ and $\varepsilon$ is shown in Fig.~\ref{t_heat_map}. This map combines Figs.~\ref{t_fixed_epsilon}~and~\ref{t_fixed_vd} and resumes the expected perfomance levels for $\varepsilon$ or $v_d$ simultaneously. 

Fig.~\ref{t_heat_map} presents a wide view on how dyads affect the evacuation dynamics. The worst situation occurs among very close \textit{colleagues} attaining desired velocities of $v_d=4\,$m/s.  This is in agreement with Ref.~\citep{FrankDorso}, where the authors showed that \textit{colleagues} persist moving together among the crowd, and thus, are responsible for slowing down the evacuation.  This phenomenon appears somehow blurred in Fig.~\ref{t_heat_map} for extremely high stressing situations (say, ${v_d>5\,}$m/s).

\subsection{Delay analysis}\label{section:delays}

\begin{figure}[ht]\centering
 \begin{subfigure}{.45\textwidth} \centering
 \includegraphics[width=\linewidth]{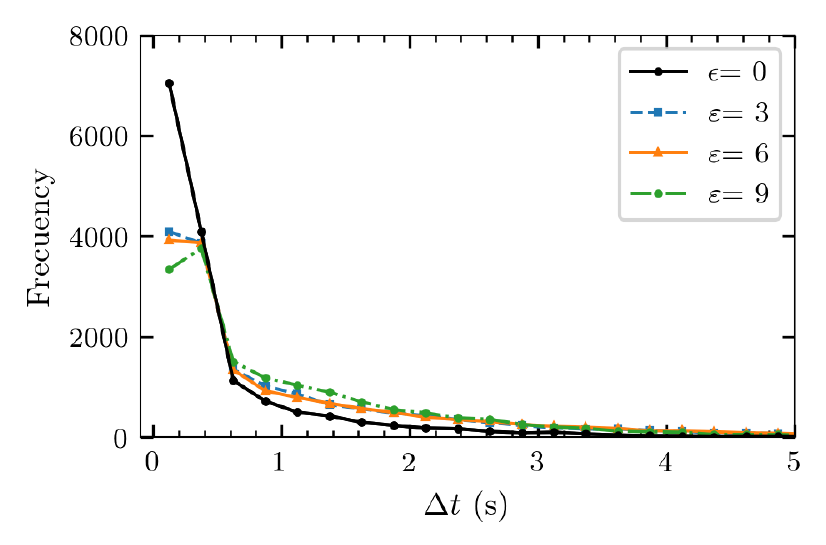}
 \caption{$v_d=3$~m/s}
 \end{subfigure}
 \begin{subfigure}{.45\textwidth} \centering
 \includegraphics[width=\linewidth]{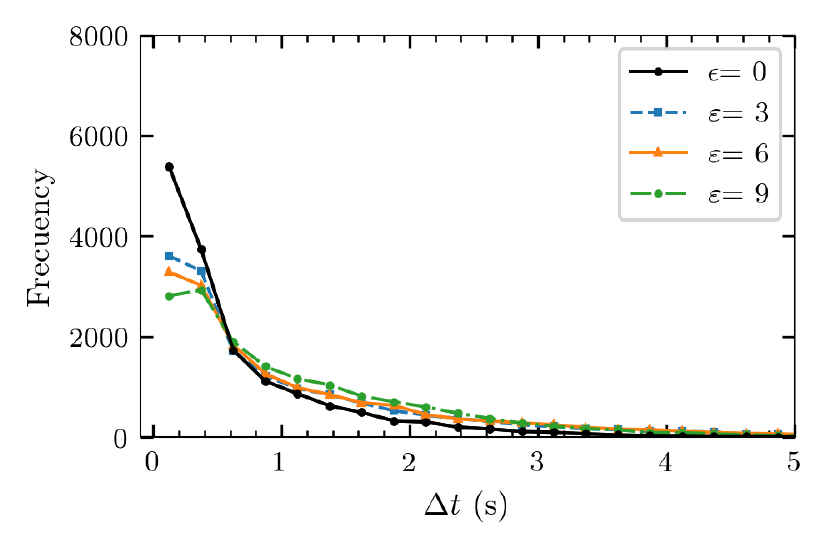}
 \caption{$v_d=5$~m/s}
 \end{subfigure}
\caption{(Color online) Occurrence distribution of delays between successive individuals. The bin size is 0.25~s. Note that the delays of all 100 simulations included. The simulation conditions were the same as in Fig.~\ref{t_fixed_epsilon} for 100\% dyads. (a) $v_d=3$~m/s. (b) $v_d=5$~m/s.}
\label{delays_ind}
\end{figure}

In order to analyze the microscopic dynamics behind the CIS and CIF regimes, we measured the time intervals between successive individuals that leave the room (regardless if they belong to the same dyad or not). We will refer to these time intervals as “delays”. Fig.~\ref{delays_ind} shows the corresponding histograms for different values of $\varepsilon$ and two representative $v_d$'s (see caption for details). Only the results for 100\% dyads are shown for clarity reasons.

\begin{figure}[ht]\centering
 \begin{subfigure}{.49\textwidth} \centering
 \includegraphics[width=\linewidth]{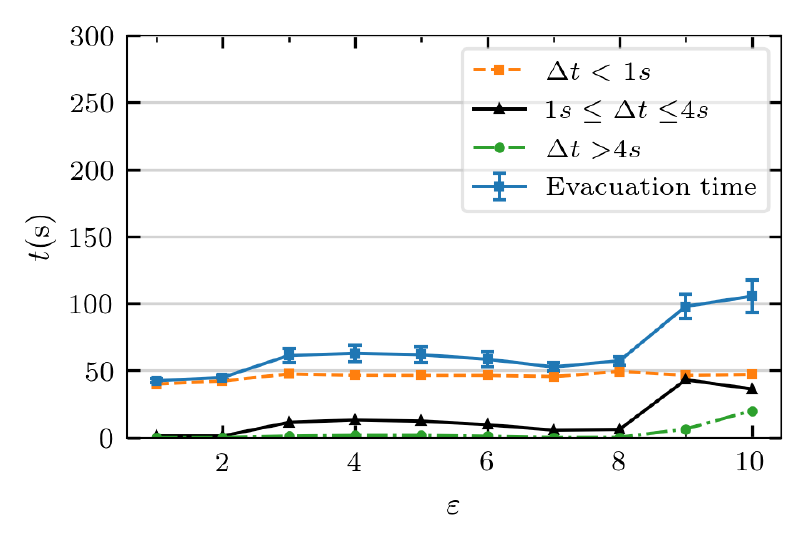}
 \caption{$v_d=1.25$~m/s}
 \end{subfigure}
 \begin{subfigure}{.49\textwidth} \centering
 \includegraphics[width=\linewidth]{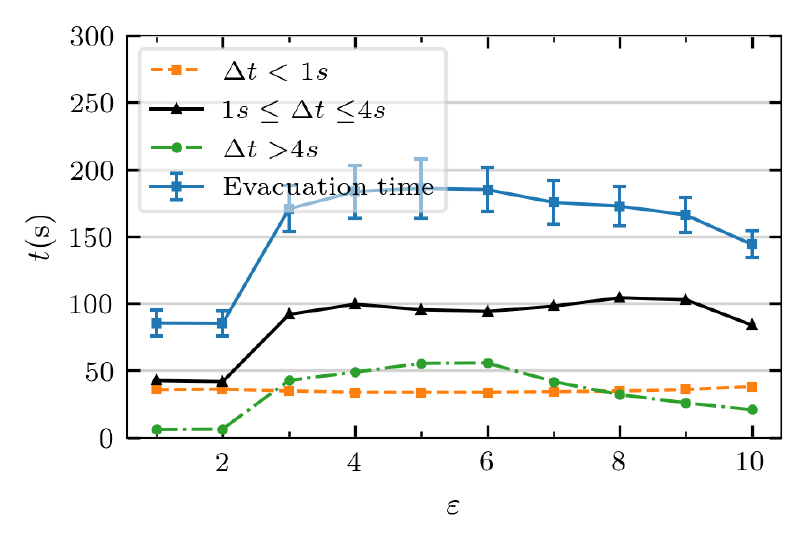}
 \caption{$v_d=3$~m/s}
 \end{subfigure}
 \begin{subfigure}{.49\textwidth} \centering
 \includegraphics[width=\linewidth]{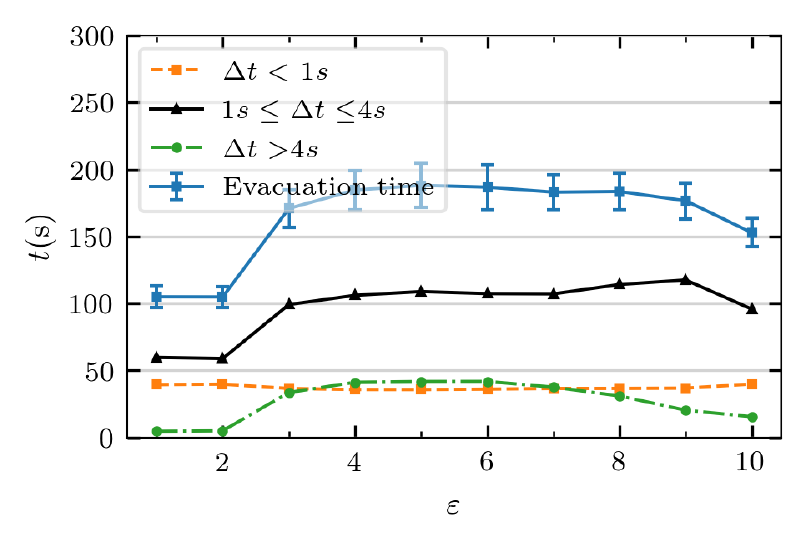}
 \caption{$v_d=5$~m/s}
 \end{subfigure}
 \begin{subfigure}{.49\textwidth} \centering
 \includegraphics[width=\linewidth]{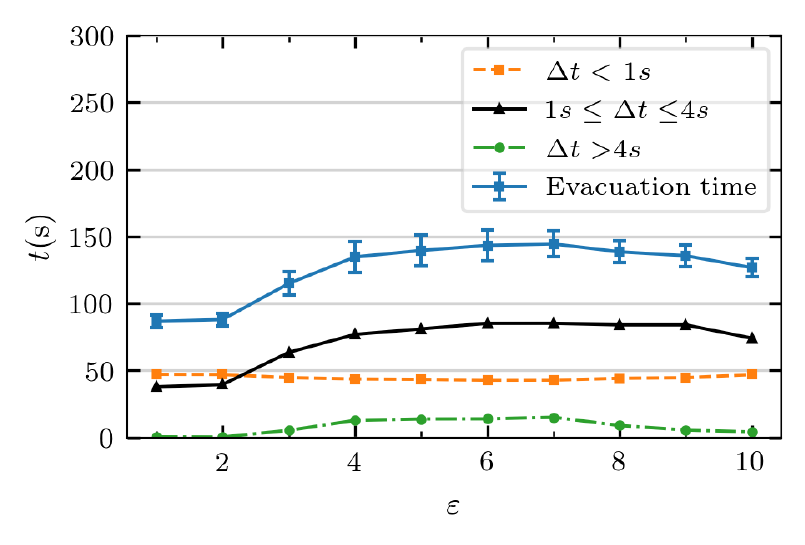}
 \caption{$v_d=8$~m/s}
 \end{subfigure}
\caption{(Color online only) Weighted sum of short, intermediate and long delays (see text for details). The total evacuation time is also included for comparison. \textbf{(a)} $v_d=1.25$~m/s. \textbf{(b)} $v_d=3$~m/s. \textbf{(c)} $v_d=5$~m/s. \textbf{(d)} $v_d=8$~m/s.}
\label{delays_sum}
\end{figure}

We can distinguish two frequency patterns in Fig.~\ref{delays_ind}. The “short delays”  (say, $\Delta t < 1\,$s) are quite considerable. The “intermediate delays” ($1\,$s${\,\leq\Delta t\leq 4\,}$s), instead, extend as a long tail. However, the situation for $v_d=5\,$m/s exhibits a heavier tail than the one for  $v_d=3\,$m/s. As a first instance, this partially explains the egress time patterns shown in Figs.~\ref{t_fixed_epsilon},~\ref{t_fixed_vd} and~\ref{t_heat_map}, but a more meaningful comparison should be made according to the weighted distribution
\begin{equation}\displaystyle
F(C)=\sum_{\Delta t_i\, \in\,C} \Delta t_i \cdot f(\Delta t_i)
\end{equation}
\noindent where $f(\Delta t_i)$ is the frequency of the delays with duration $\Delta t_i$ and $C$ means any category set. For simplicity we will analyze only three categories of delays: short ($\Delta t< 1$~s), intermediate (1~s$\leq \Delta t\leq 4$~s) and long ($\Delta t > 4$~s). Fig.~\ref{delays_sum} shows each distribution as a function of $\varepsilon$. 

Recall that $v_d=1.25\,$m/s is out of the panic scenario, and thus, we will not attempt to analyze this case. The reader can find some notes on this issue in \ref{appendix:lowvdhigheps}.

Notice that short delays remain unchanged for every value of $\varepsilon$ and any given $v_d$. Thus, we infer that the CIS or CIF regimes should be related to the presence of intermediate and long delays. Fig.~\ref{delays_sum} also shows that the intermediate delays become relevant in context of the panic. These contribute along both the CIS and CIF regimes. The long delays contribute in less degree and diminish along the CIF regime (say, for $\varepsilon>6$).

We further examined the animations of the simulations, and observed that individuals get released from the blocking clusters more frequently in the CIF regime, and consequently, the time that the exit remains blocked diminishes. This explains the decrease of the long delays and the overall decrease of the egress time. The observed behaviors relating the time lapses and the blocking clusters were also reported in Ref.~\cite{cornes2020}.

\begin{figure}[ht]\centering
  \begin{subfigure}{.49\textwidth} \centering
 \includegraphics[width=\linewidth]{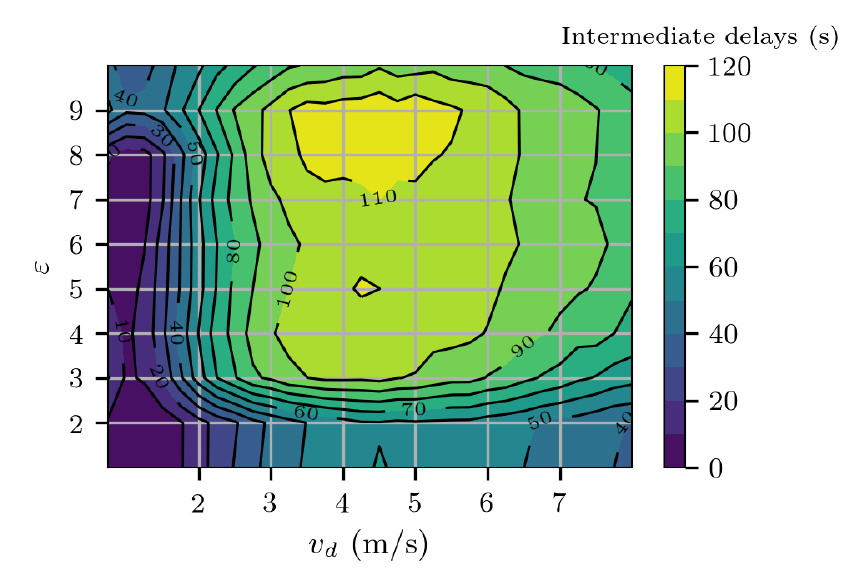}
 \caption{Sum of intermediate delays}
 \end{subfigure}
 \begin{subfigure}{.49\textwidth} \centering
 \includegraphics[width=\linewidth]{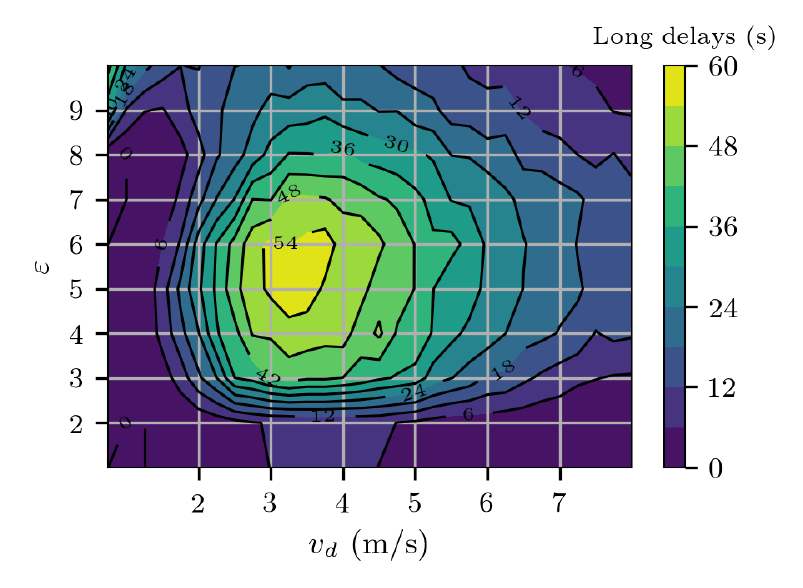}
 \caption{Sum of long delays}
 \end{subfigure}
\caption{(Color online only) Heat-maps for the weighted the sum of \textbf{(a)} intermediate delays (1~s $\leq\!\Delta t\!\leq$~4~s) and \textbf{(b)} long delays ($\Delta t\!>$~4~s). The simulation conditions are the same as in Fig.~\ref{t_fixed_epsilon}.}
\label{delays_heatmap}
\end{figure}

Fig.~\ref{delays_heatmap} summarizes the above results as a function of $v_d$ and $\varepsilon$. We can see that the intermediate delays attain a maximum at the top of the heat-map ($8<\varepsilon<9$, that is, for intimate couples), while this maximum moves to  the middle of the map ($5<\varepsilon<6$) for long delays. We conclude that there is a “displacement” from long to intermediate ones, which ultimately implies an improvement for the evacuation time. Finally, for $\varepsilon>9$, both categories of delays decrease.

\FloatBarrier
\subsection{Ordered vs. disordered dynamics in the bottleneck } 
\label{section:intragroup}

We now open the question on how many of the individuals that egress consecutively are actually dyad-partners. This question concerns the “ordering” of the evacuation process, where by “order” we mean that dyad members egress together. For instance, in an “ideally ordered” evacuation process where 100\% of the individuals are grouped, exactly 50\% of the delays should correspond to partners | the first one egresses, the second one is his (her) dyad-partner, then the third one is from another dyad, the fourth one is his (her) partner and so on. 

\begin{figure}[th]\centering
\begin{subfigure}{.49\textwidth} \centering
 \includegraphics[width=\linewidth]{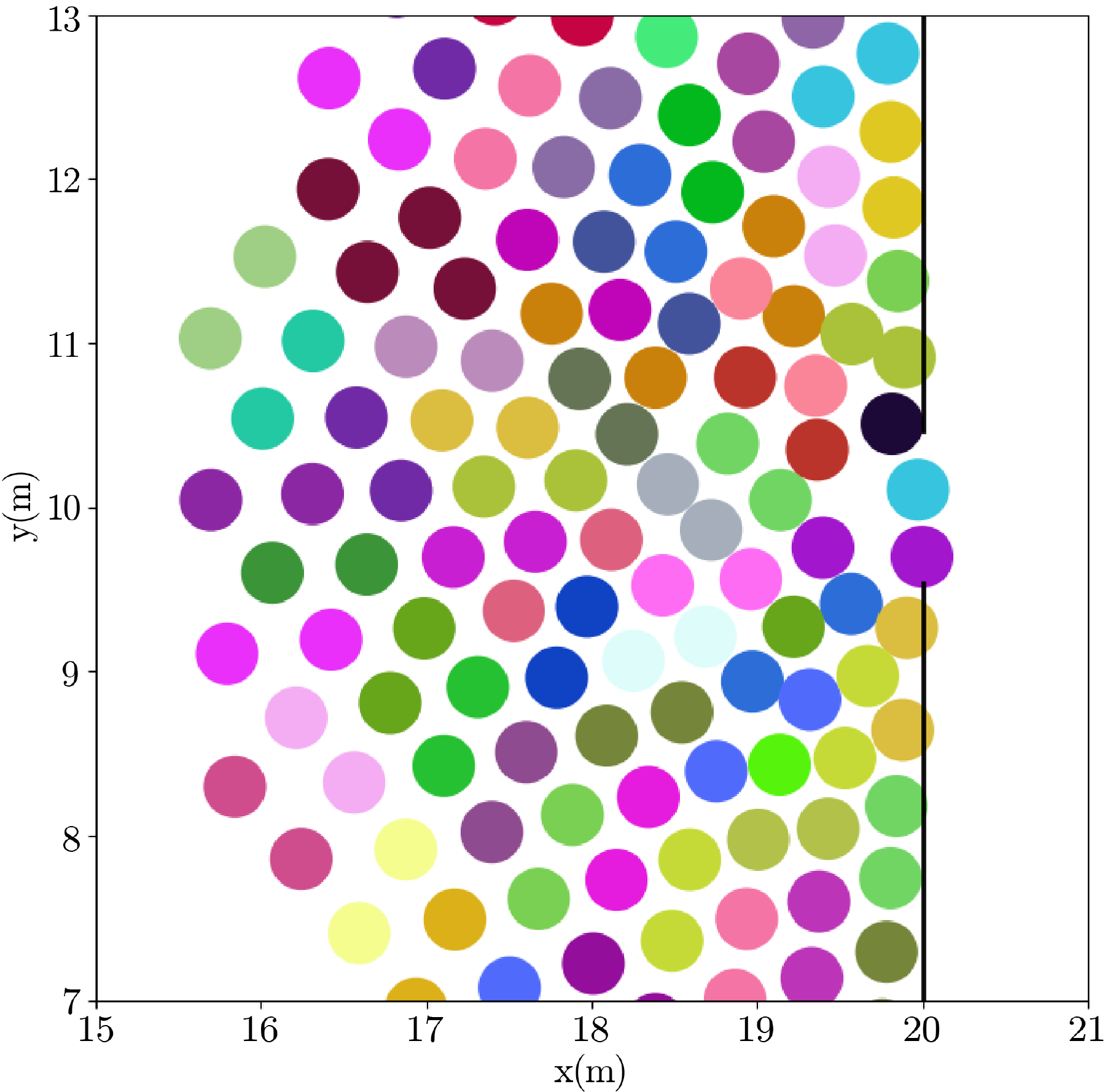}
 \caption{$t=68.10$~s}
 \label{dyad_interrumpted_a}
 \end{subfigure}
\begin{subfigure}{.49\textwidth} \centering
\includegraphics[width=\linewidth]{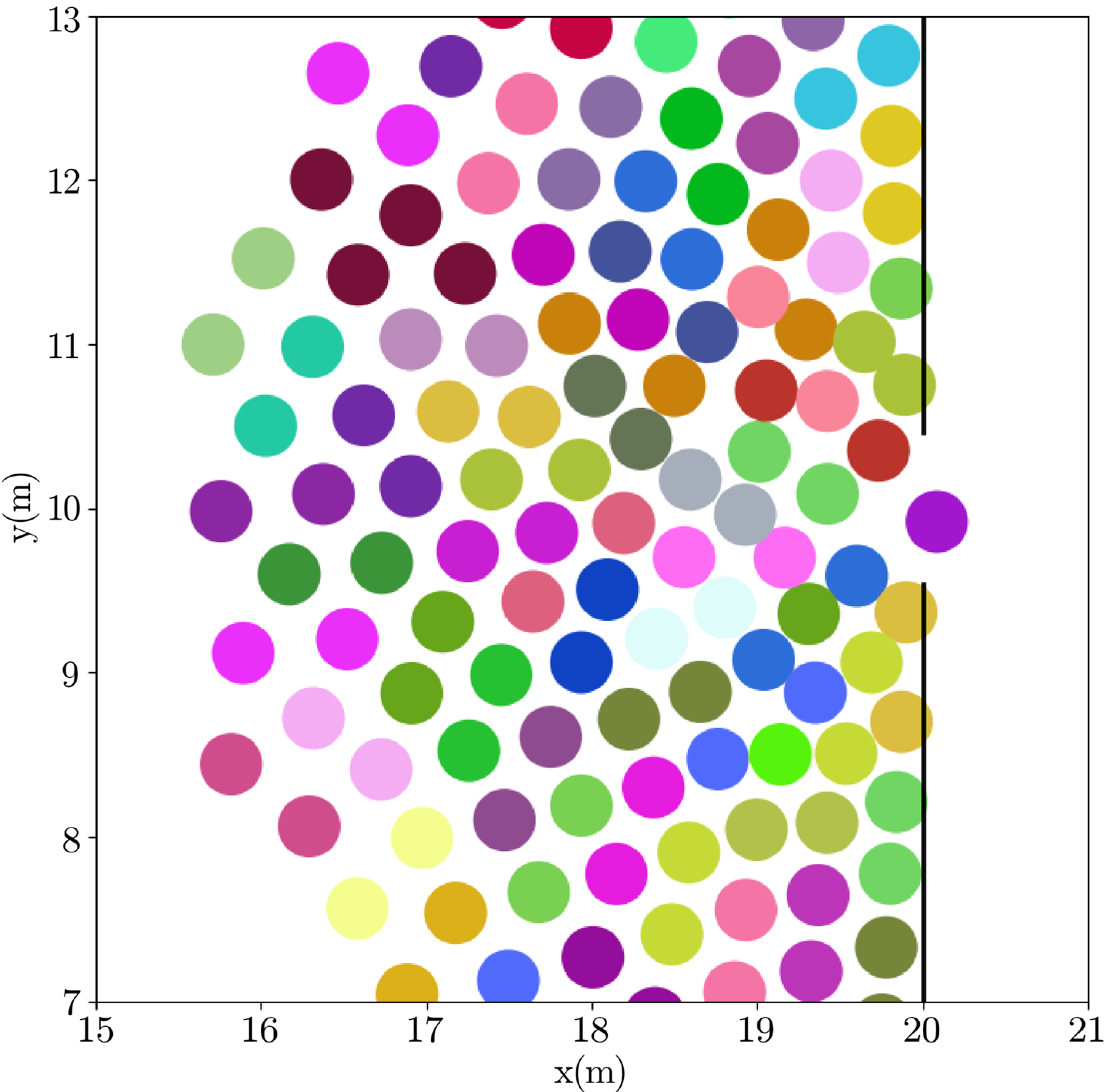}
 \caption{$t=69.95$~s}
 \end{subfigure}  
 \begin{subfigure}{.6\textwidth} \centering
\includegraphics[width=\linewidth]{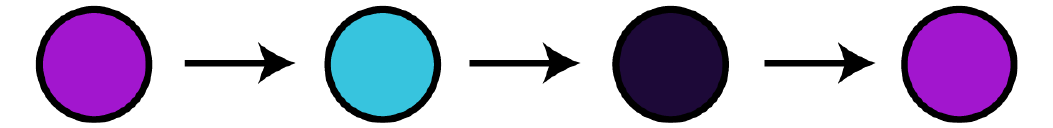}
 \caption{Order of egress}
 \label{dyad_interrumpted_c}
 \end{subfigure}
 \caption{(Color online only) Snapshots of an evacuation process for $v_d=5$~m/s and $\varepsilon=5$. Each dyad is represented by a different color (online version only). The \textit{purple} individual egresses from the bottom half of the door at $t=68.10$~s. Then the \textit{light blue} and \textit{dark blue} individuals egress from the top half of the door at $t=68.15$~s and $t=68.55$~s, respectively. However, as the second member of the \textit{purple} dyad egresses from the bottom half at $t=69.95$~s, we consider that the dyad partners do not miss each other, and both egress (seemingly) together. }
\label{dyad_interrumpted}
\end{figure}

Notice that as the door is wide enough for two pedestrians to exit simultaneously, there is a chance that an individual might sneak in between a dyad, without loosing the attractive feelings between the partners (say, without missing each other). An example of such a situation is shown in Fig.~\ref{dyad_interrumpted} (see caption for details). We will, therefore, consider that both dyad members egress together if not more than three individuals evacuated between them, as sketched in Fig.~\ref{dyad_interrumpted_c}.

\FloatBarrier

\begin{figure}[ht]\centering
 \includegraphics[width=0.7\linewidth]{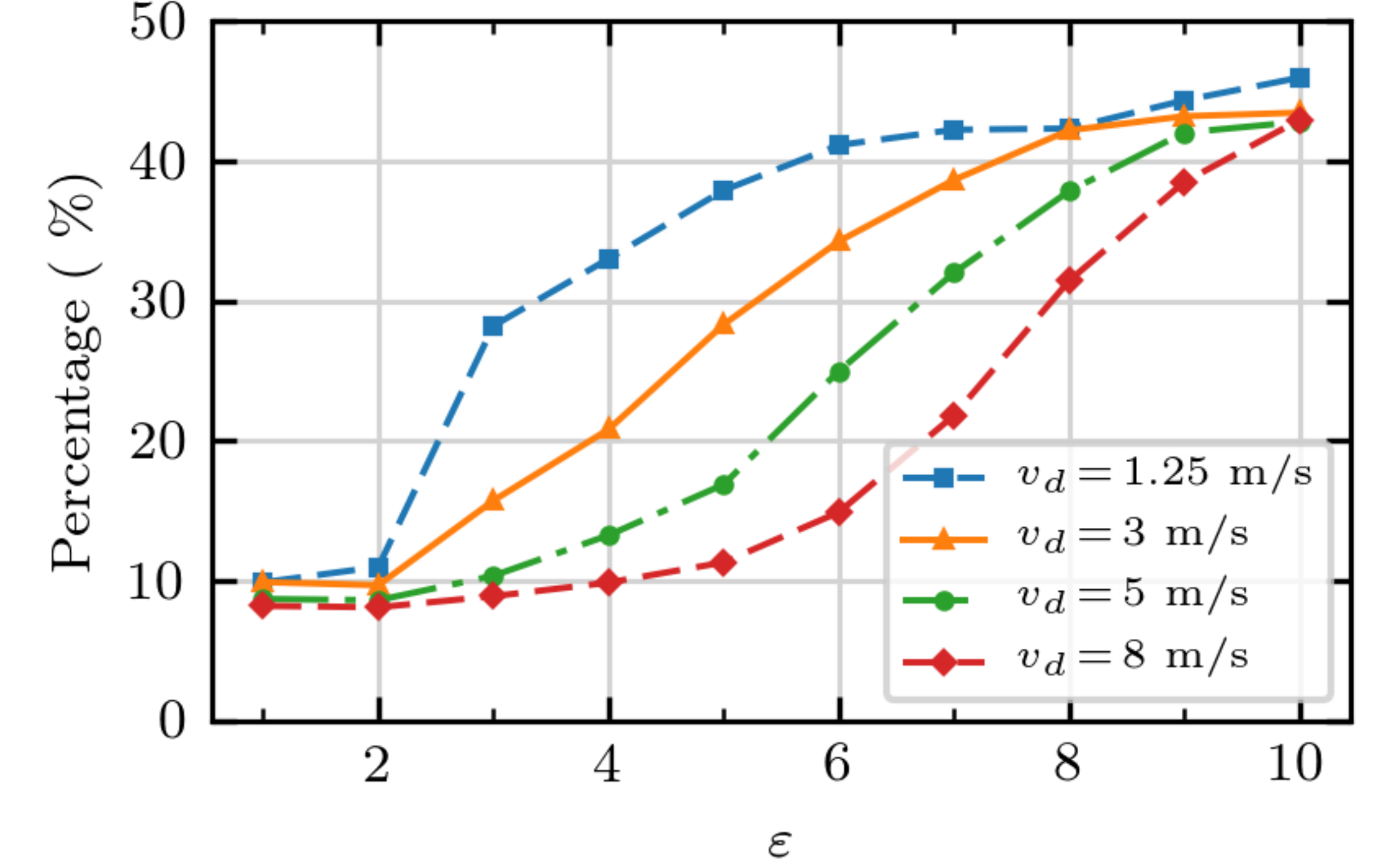}
 \caption{(Color online) Percentage of individuals that egress together with their partners. The “ideally” ordered scenario corresponds to 50\%. The simulation conditions are the same as in Fig.~\ref{t_fixed_epsilon}.}
 \label{delays_percentage}
\end{figure}

Fig.~\ref{delays_percentage} shows the fraction of individuals that egress together with their dyad-partner. 
As $\varepsilon$ increases for any fixed value of $v_d$, the number of partners that egress together increases, which was to be expected, since stronger attractive feelings will motivate partners to remain coupled. In particular, when $\varepsilon=10$ this fraction surpasses 40\% for all the explored anxiety levels, almost reaching an “ideally ordered” scenario.

On the other hand, for any fixed value of $\varepsilon$, as $v_d$ increases the ordering fraction decreases. This means that the dyads miss the egress “order” as their anxiety level increases (say, $v_d$ increases). This is quite noticeable within the range of \textit{colleagues} and \textit{couples} ($4<\varepsilon<6$). \textit{Intimate couples}, however, are scarcely affected by panic (within the explored range of $v_d$).

\begin{figure}[bht]\centering
 \includegraphics[width=0.55\linewidth]{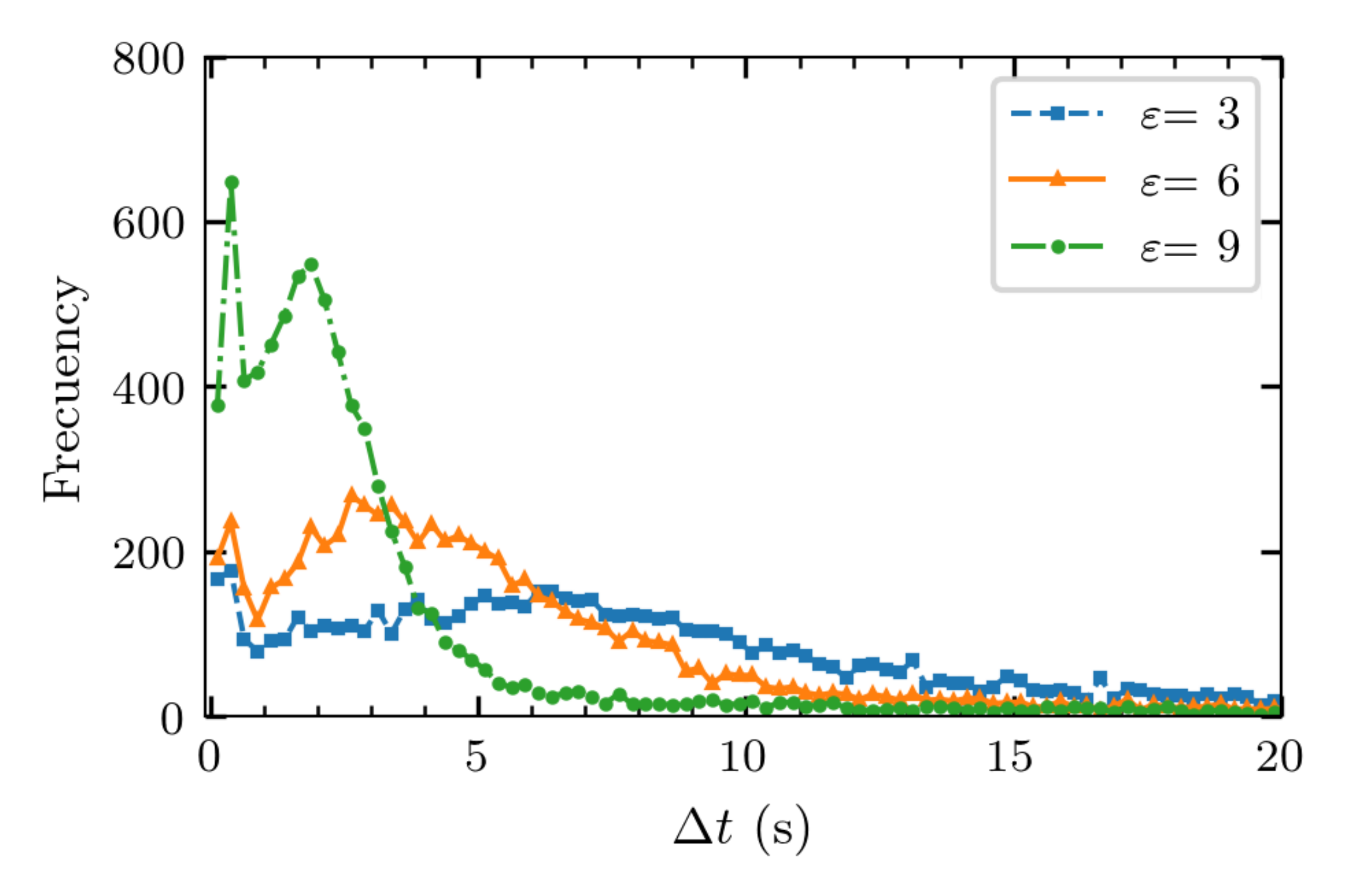}
\caption{(Color online only) Intragroup delay distribution for three representative values of $\varepsilon$ for $v_d=5\,$m/s. All the individuals where grouped in dyads. The bin size of the histograms is 0.25$\,$s.  }
\label{delays_dyads}
\end{figure}

We further linked the degree of “order” in the evacuation process with the corresponding egress delays. Fig.~\ref{delays_dyads} shows the distribution of delays within individuals belonging to the same dyad (see caption for details). We call these “intra-group delays”.  

\begin{figure}[ht]\centering

 \begin{subfigure}{.49\textwidth} \centering
 \includegraphics[width=\linewidth]{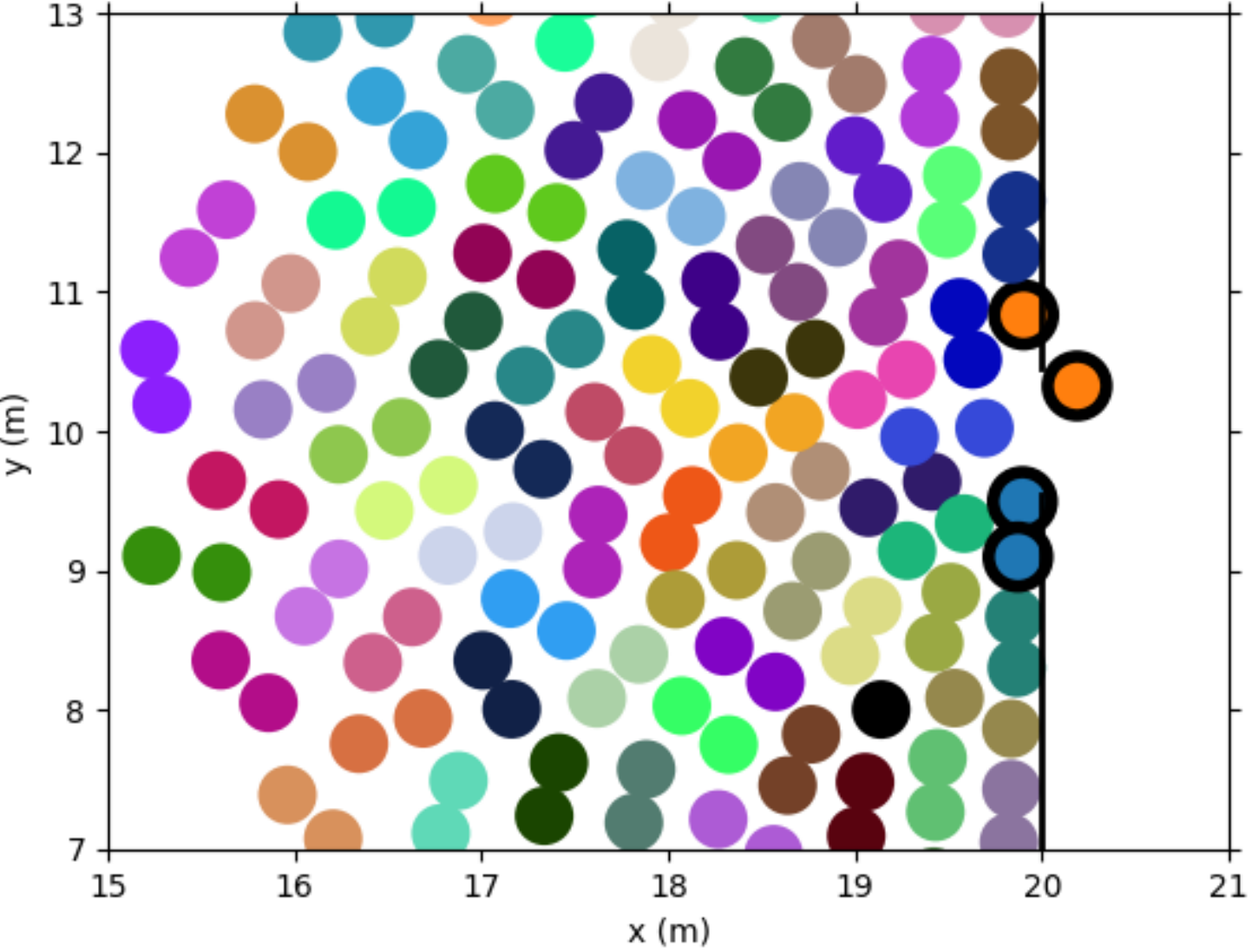}
 \caption{$t=21.15$~s}
 \end{subfigure} 
 \begin{subfigure}{.49\textwidth} \centering
 \includegraphics[width=\linewidth]{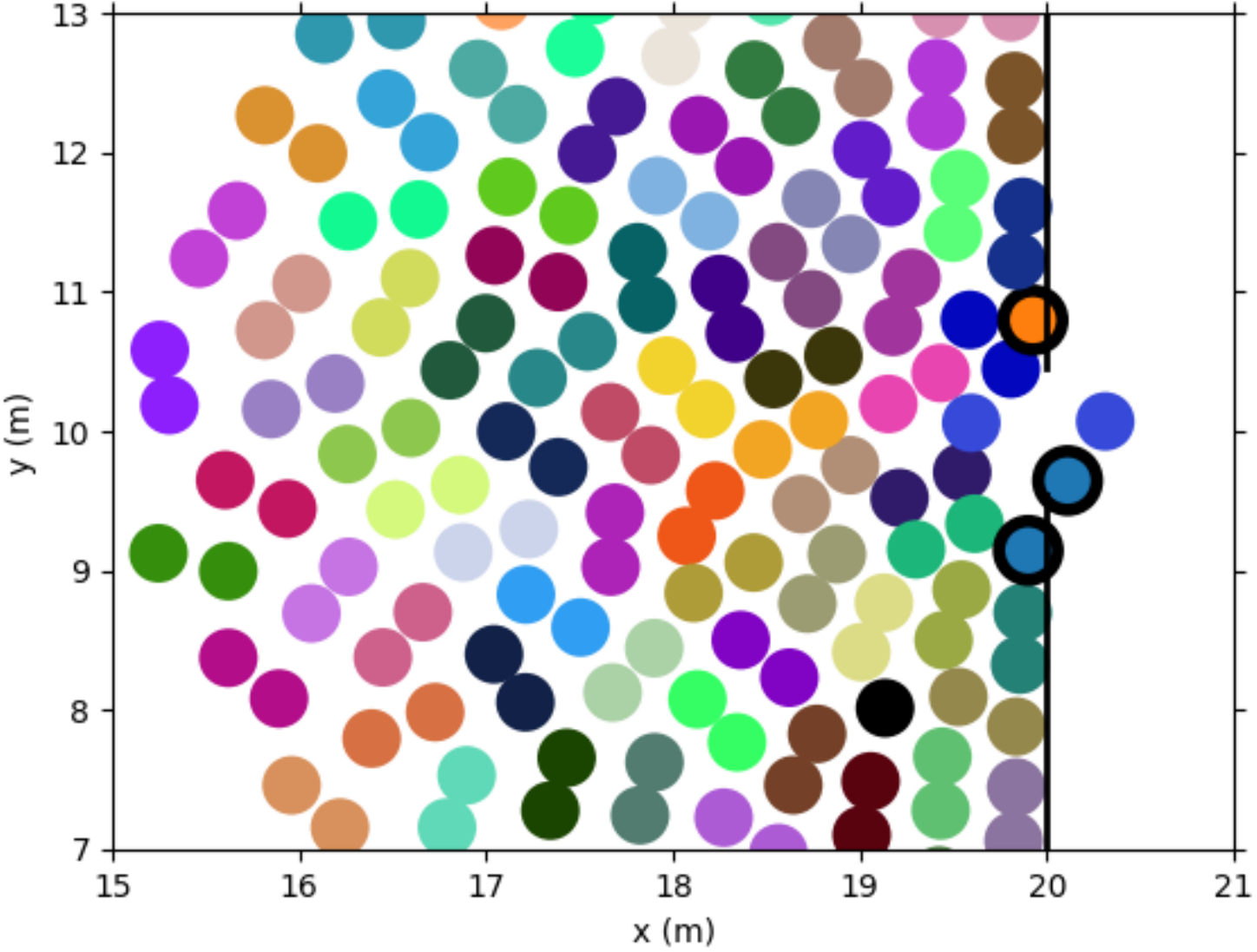}
 \caption{$t=22.60$~s}
 \end{subfigure} 
 \begin{subfigure}{.49\textwidth} \centering
 \includegraphics[width=\linewidth]{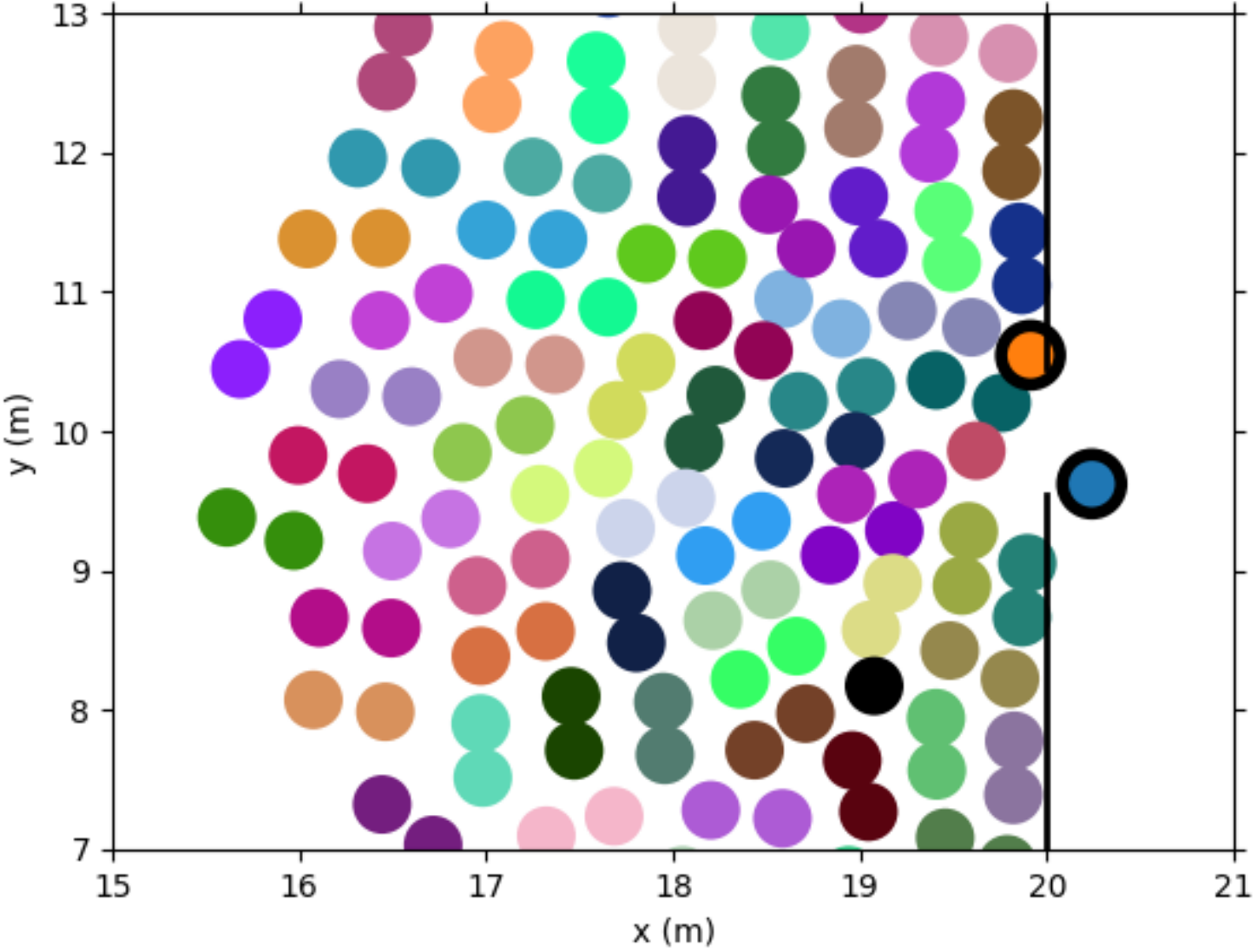}
 \caption{$t=59.75$~s}
 \end{subfigure} 
 \begin{subfigure}{.49\textwidth} \centering
 \includegraphics[width=\linewidth]{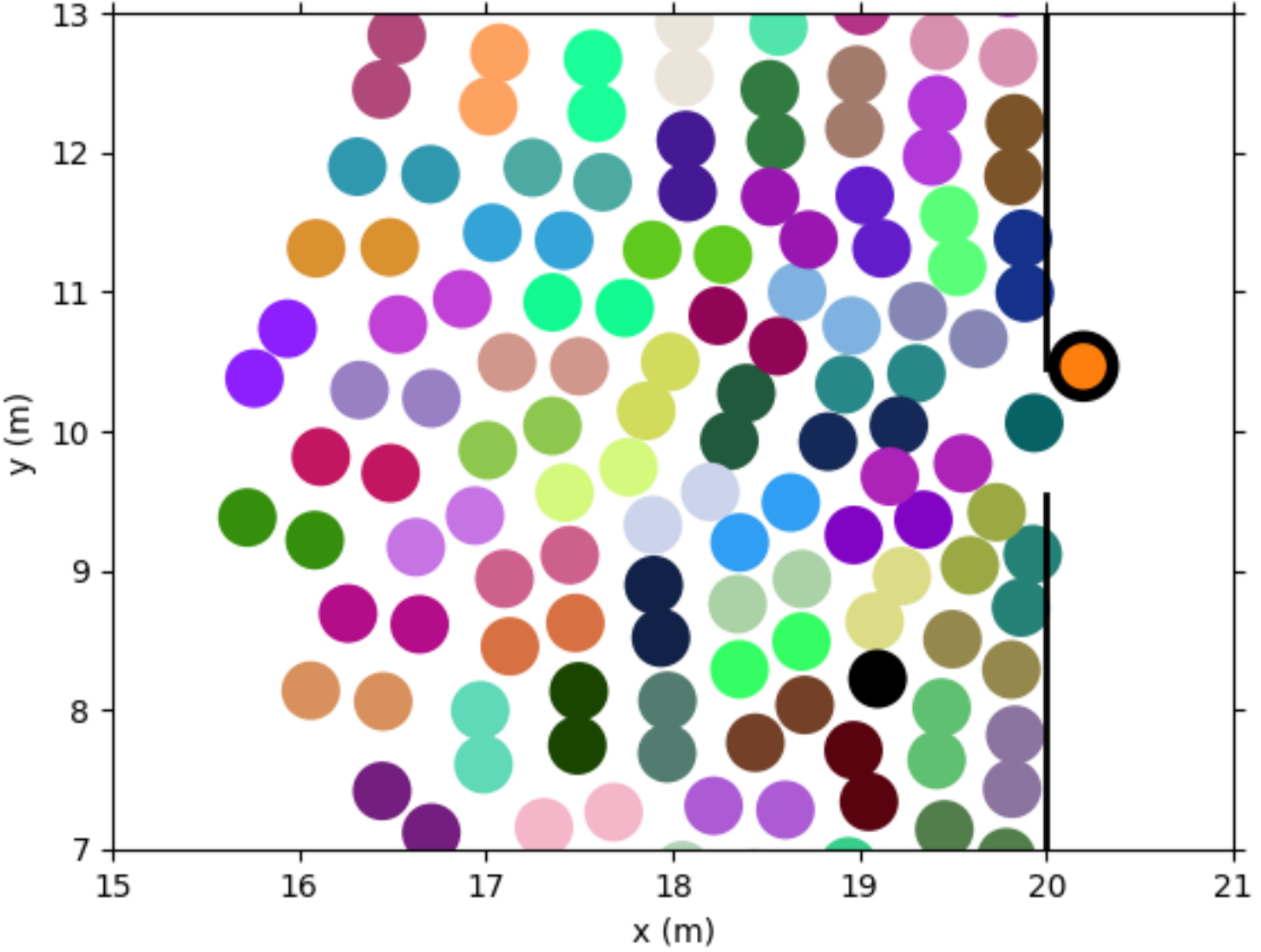}
 \caption{$t=61.65$~s}
 \end{subfigure}
 \caption{(Color online only) Snapshots of an evacuation process for 100\% dyads and $v_d=5$~m/s. The door is centered at $x=20$~m, $y=10$~m and has a 0.92~m width. All dyads have a very strong attractive feelings ($\varepsilon=9$). Each dyad is represented by a different color, and those with a black border  have a fairly long intragroup delay (online version only). The \textit{orange} partners egress at snapshots (a) and (d), while the \textit{blue} partners egress at  snapshots (b) and (c), respectively.  The intra-group delay for the \textit{orange} dyad is 40.5$\,$s, while the corresponding delay for the \textit{blue} one is 37$\,$s. }
\label{broken_links}
\end{figure}

It is clear from Fig.~\ref{delays_dyads} that the more intense attractive feelings, the shorter the intra-group delays. For instance, the intra-group delays that are longer than 15$\,$s represent 22\% for $\varepsilon=3$, 12\% for $\varepsilon=6$ and 6\% for  $\varepsilon=9$. These occur often on the sides of the door, as illustrated in Fig.~\ref{broken_links} (see caption for details).

We conclude from this Section that “ordering” is a major issue whenever social groups are present. Strong feelings within the couples ensures that both will evacuate together. On the contrary, weak feelings will allow individuals to sneak in between the dyads, making partners miss each other, and therefore slowing down the evacuation.

\FloatBarrier

\subsection{Ordered vs. disordered dynamics in the corridor}\label{section:results_fd}

We report in this Section the dynamical consequences of introducing social groups in a corridor (within the context of the SFM). We focus on the fundamental diagram (\textit{i.e.} flux vs. density, see Section ~\ref{section:fd_intro}) for crowds where 70\% of the individuals belong to groups of up to 5 individuals. The sampling procedure is as shown in Fig.~\ref{fd_region}. 

\begin{figure}[th]\centering
\begin{subfigure}{.49\textwidth}\centering
 \includegraphics[width=1.06\linewidth]{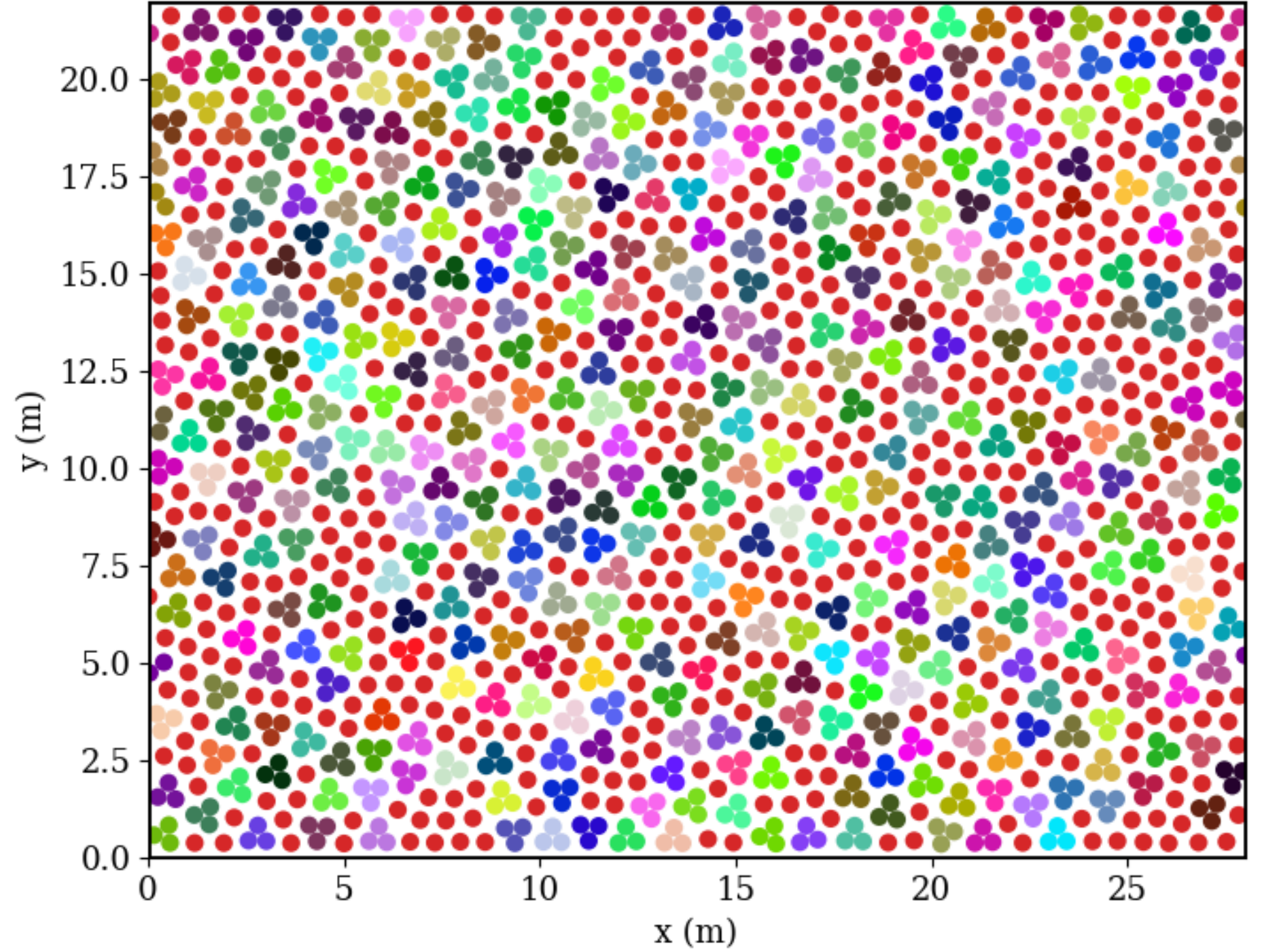}
 \caption{Groups of 3 members}
 \end{subfigure}
\begin{subfigure}{.49\textwidth} \centering
\includegraphics[width=0.935\linewidth]{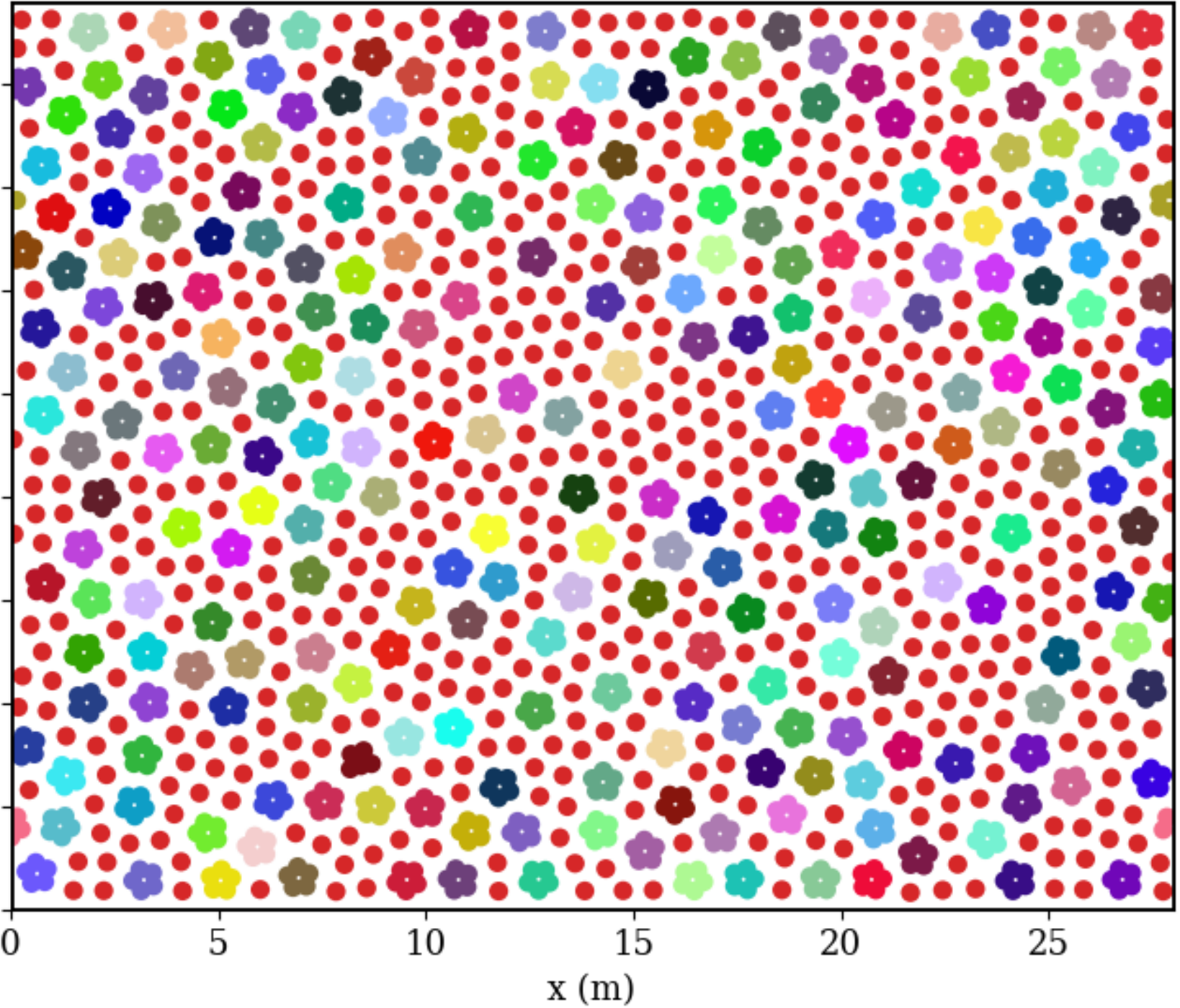}
 \caption{Groups of 5 members}
 \end{subfigure}
 \caption{(Color online only) Snapshots of simulated processes for corridors with groups of 3 and 5 members, respectively, attaining strong attractive feelings ($\varepsilon=8$). Each group is represented by a different color, and individuals that are not members of a group are represented in red. In both cases, the total number of individuals is $N=1848$, which divided by the area of the corridor gives the global density $\rho=3$~P/m$^2$ (this corresponds to the the free-flow regime). The desired velocity of all individuals was $v_d=1\,$m/s. Both situations are stationary, right after the pedestrians are accommodated within the corridor.}
\label{big_individuals}
\end{figure}

Fig.~\ref{big_individuals} captures how individuals are grouped in our model while the corridor is in a non-congested regime (see caption for details).

\FloatBarrier

\begin{figure}[ht]\centering
 \includegraphics[width=0.75\linewidth]{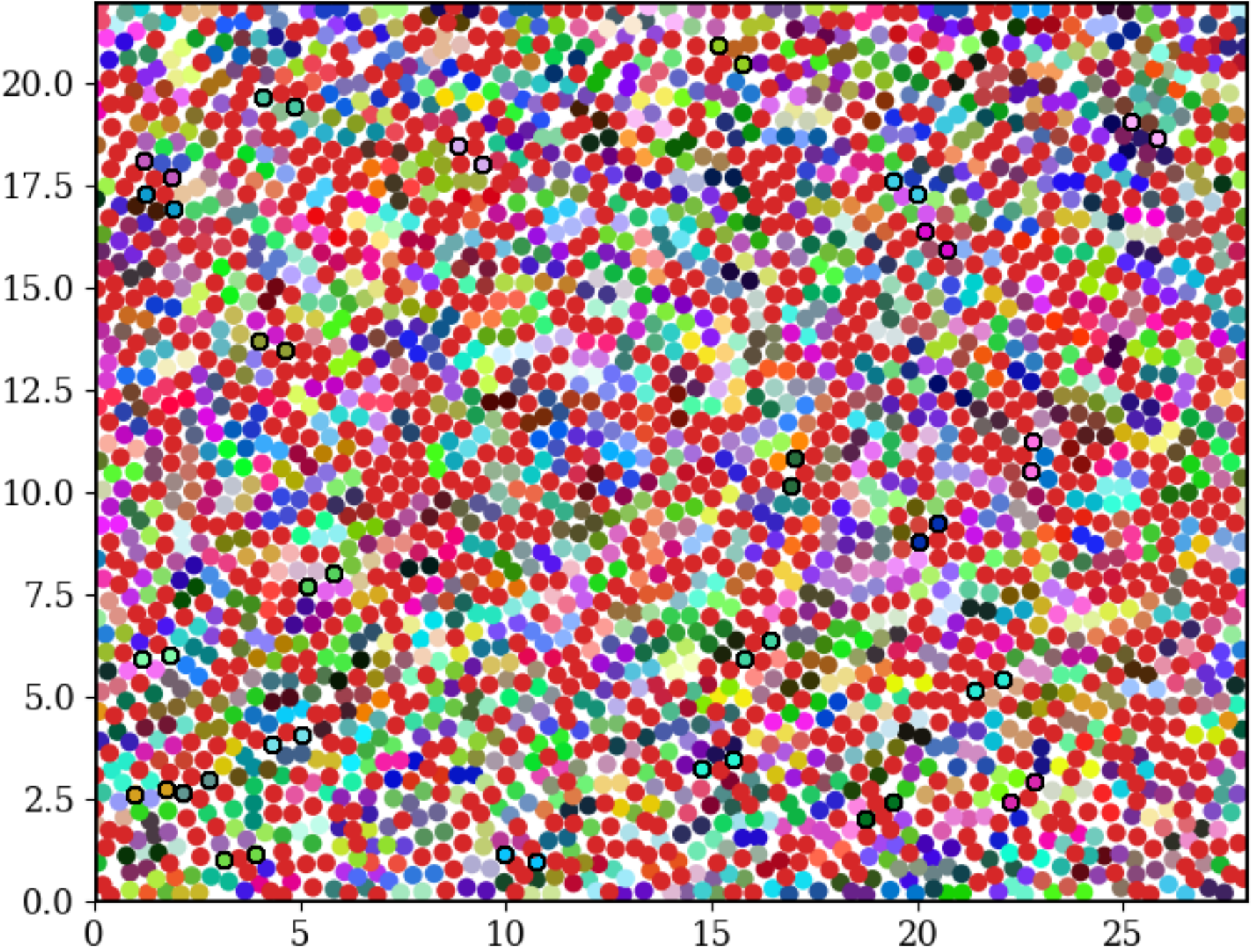}
 \caption{(Color online only) Snapshot at $t=32.2\,$s of a corridor with dyads attaining $\varepsilon=5$. Each group is represented by a different color, and individuals that are not members of a group are represented in red. The global density is $\rho=5$~P/m$^2$ (which corresponds to the congested regime). The desired velocity for all the individuals was $v_d=1\,$m/s. The dyads with a black border are a few examples of how other individuals can sneak in between groups at high densities.}
\label{congested_corridor}
\end{figure}

Recall that “ordering” in Section~\ref{section:intragroup} means that groups egress together, while no one can sneak in between. The situation in the corridor is somehow different, although some individuals might sneak in between the members of a group when the attractive feelings are those associated to \textit{colleagues} and \textit{couples} (say, $\varepsilon<7$) and the corridor is congested. This situation is shown in Fig.~\ref{congested_corridor}. We will therefore associate these inhomogeneities to a somewhat “disordered” crowd.

\FloatBarrier

\begin{figure}[ht]\centering
 \includegraphics[width=0.7\linewidth]{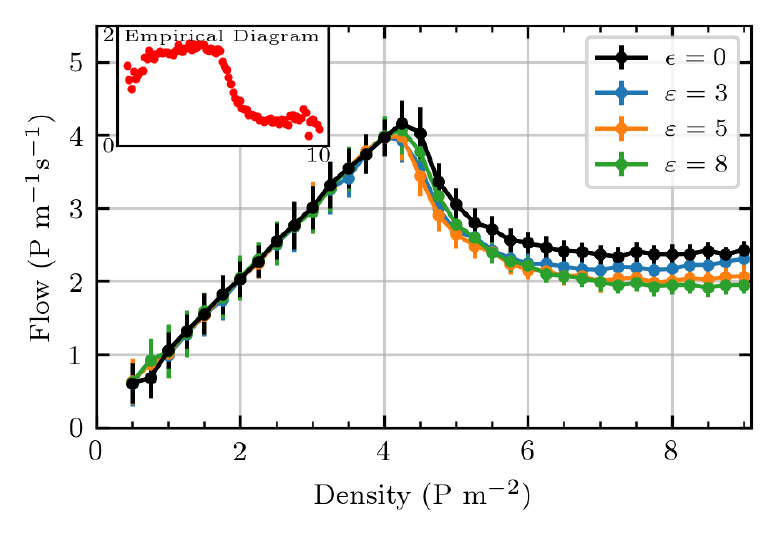}
 \caption{(Color online only) Mean flow as a function of the global density ($\rho=N/\mathrm{Area}$) for the corridor of 22~m wide and 28~m long (with periodic boundary conditions). The flow was sampled within a circle of $R=1\,$m in the middle of the corridor. The data points are mean values over 80$\,$s, sampled at a frequency of 2 times per second. 70\% of the individuals were grouped in dyads, and every dyad has the same attractive intensity $\varepsilon$ in each simulation. The desired velocity was $v_d=1\,$m/s for all the individuals. The empirical diagram from Ref.~\cite{helbing_FD} is shown in the inset.}
 \label{fd_dyads}
\end{figure}

Fig.~\ref{fd_dyads} shows the fundamental diagram for 70\% of the pedestrians grouped in dyads. A slight slowing down of the flux can be observed in the congested regime (say, $\rho> 4.5$~P/m$^2$) as $\varepsilon$ increases. Notice that varying $\varepsilon$ does not significantly change the behavior of the flow (within the explored range). We present in \ref{appendix:vel_profile} a more detailed analysis regarding the slight differences between the curves in the congested regime, by computing the velocity profiles among the corridor. 

\begin{figure}[bht]\centering
 \includegraphics[width=0.7\linewidth]{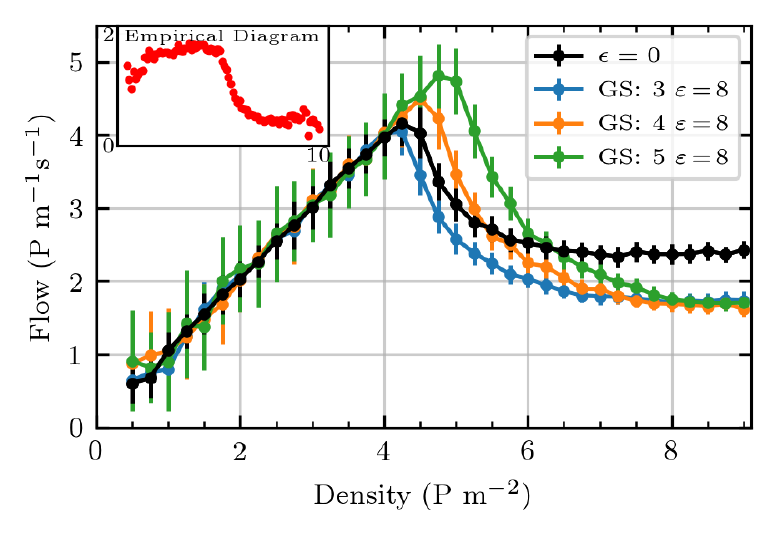}
 \caption{(Color online only) Fundamental diagram for a crowd moving along a corridor, where 70\% of the pedestrians belong to groups of up to 5 individuals. The attractive intensity was $\varepsilon=8$. GS stands for the group size (all groups were of the same size in each simulated process). The black curve for $\epsilon=0$ corresponds to the no grouping situation. The desired velocity was $v_d=1\,$m/s for all individuals.  }
 \label{fd_intimate}
\end{figure} 

We hypothesize that the slight slowing down could become more relevant when groups are enlarged (say, of up to 5 members). Fig.~\ref{fd_intimate} shows the fundamental diagram obtained for groups with 3 to 5 members with an attractive intensity of $\varepsilon=8$. We can see that the free-flow regime (say, the non-congested regime) extends to higher densities as the number of group members increases. This means that groups move in a free-flow environment at densities up to $4-5\,$P/m$^2$.

A re-examination of Fig.~\ref{big_individuals} shows that there is more space between groups as they become larger. Thus, we presume that the “effective density” of the crowd is somewhat lower than the global density (say, $N$ divided by the total corridor area). We will  discuss this point in Section~\ref{section:discussion}.

\begin{figure}[bht]\centering
 \includegraphics[width=0.7\linewidth]{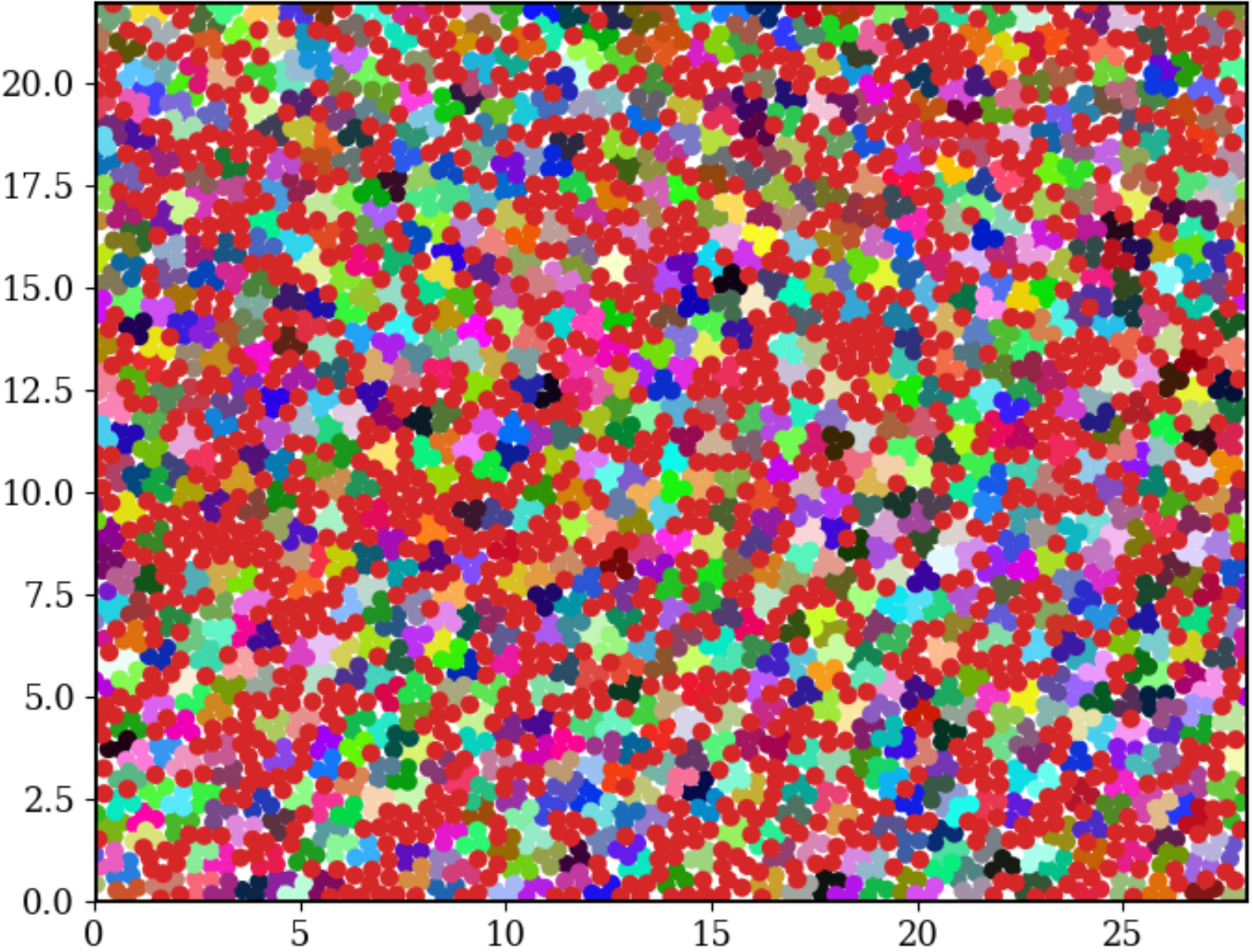}
 \caption{(Color online only) Snapshot at $t=33.5\,$s for a corridor with groups of 3 members attaining $\varepsilon=8$. The global density is $\rho=7\,$P/m$^2$ (which belongs to the congested regime). The simulation conditions are otherwise the same as in Fig.~\ref{big_individuals}. }
 \label{congested_corridor_gs5}
\end{figure}

Fig.~\ref{fd_intimate} also shows that no matter the size of the groups (within the explored range), the flow slows down to a common value for extremely high global densities. This corresponds to a very compact scene, as shown in Fig.~\ref{congested_corridor_gs5} where all the pedestrians are in contact. 

The flow in Fig.~\ref{fd_dyads} drops to 2$\,$P(ms)$^{-1}$ for dyads at very high densities (and $\varepsilon=8$). The corresponding flow in Fig.~\ref{fd_intimate} drops to a somewhat lower value, no matter the group size (within the explored range). This means that the group size is not a relevant magnitude at very high densities, except for dyads. Fig.~\ref{congested_corridor_gs5} shows that more than two members per group form quite closely shaped structures. As a drawback, these groups can not move among the crowd as smoothly as dyads or single individuals do. Moreover, as groups attain stronger attractive feelings (such as $\varepsilon=8$), this drawback becomes more relevant. Thus, it seems reasonable to obtain a more significant slowing down with respect to the dyads situation (at very high densities).

We summarize the above results as follows. Social groups introduce a somewhat inhomogeneous (disordered) scene along the corridor. This yields to a fundamental diagram that differs from the one reported for single individuals. The differences, however, become more significant if the  groups' size exceeds two members. The first difference occurs at the intermediate density interval, where the free-flow regime extends into higher densities (say, $\rho=4-5\,$P/m$^2$). The second difference, though, concerns the slowing down. Social groups slow the flow down more pronouncedly than single individuals at the extremely congested regime.  However, the size of the groups seems to not play a role in this phenomenon (within the explored group sizes).

\section{Discussion} \label{section:discussion}

We analyzed in Section~\ref{section:results} the evacuation dynamics of social groups in the bottleneck and the corridor scenarios. Although quite different situations, both show that the presence of attractive feelings introduce additional delays (bottlenecks) or slowing downs (corridor) in the movement dynamics. We now discuss how these occur according to the intensity of the attraction. We focus on dyads, but the discussion can be extended to larger groups. 

\begin{figure}[ht]\centering
 \begin{subfigure}{.49\textwidth} \centering
 \includegraphics[width=\linewidth]{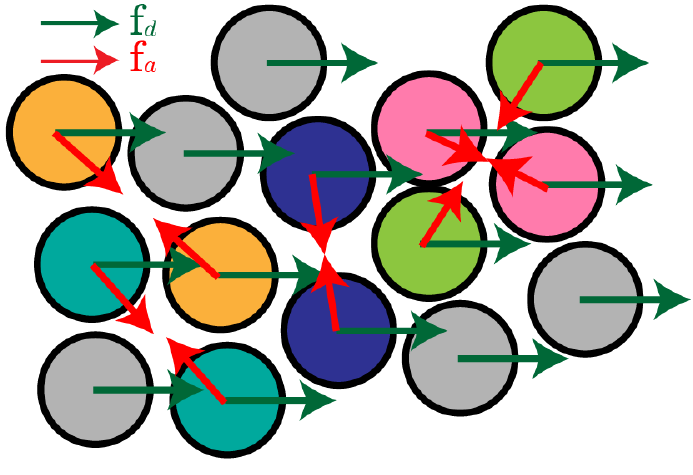}
 \caption{Weak $\varepsilon$}
 \label{crowd_scheme_a}
 \end{subfigure}
 \begin{subfigure}{.49\textwidth} \centering
 \includegraphics[width=\linewidth]{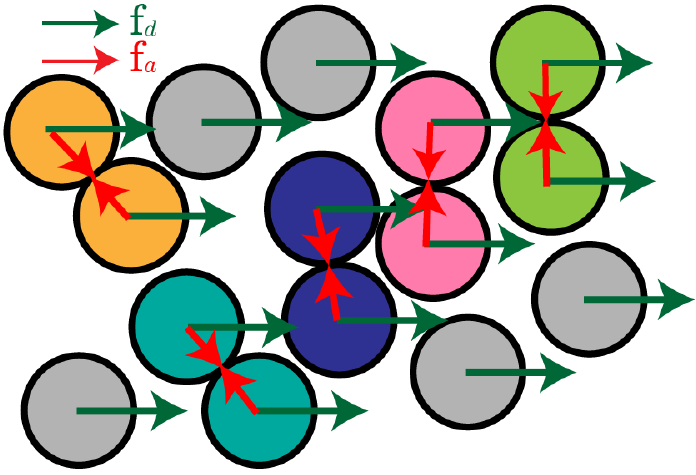}
 \caption{Strong $\varepsilon$}
 \label{crowd_scheme_b}
 \end{subfigure}
\caption{(Color online only) Schemes comparing crowds with different types of dyads. Each color represents a dyad while gray individuals are not grouped. The desired and attractive forces are shown in green (\textbf{f}$_d$) and red (\textbf{f}$_a$), respectively. All the desire forces point to the right for simplicity, but it is also possible to point to a nearby  target. \textbf{(a)} Crowd with dyads attaining weak attractive feelings intensity $\varepsilon$ (\textit{colleagues, couples}). \textbf{(b)} Crowds with dyads attaining very strong $\varepsilon$ (\textit{intimate couples}).}
\label{crowd_scheme}
\end{figure}

We present two schemes in Fig.~\ref{crowd_scheme}. The scheme on the left displays the situation of “weakly” attracted partners, where mutual feelings are not strong enough to get into body contact. The scheme on the right displays the situation expected for intimate partners attaining a “strong” attractive feeling, and thus, getting into the contact distance.

As shown in Fig.~\ref{crowd_scheme_a}, when the attractive feelings are not strong enough to guarantee that all dyads will be at contact distance, some individuals are able to squeeze among partners. This brings out negative consequences to the evacuation. For instance,  the pink  individual on the rightmost of Fig.~\ref{crowd_scheme_a} squeezed through the green dyad. But the green dyad is still blocking the way to the pink individual on the left. At this point, the whole pink dyad gets delayed since none of the pink individuals will leave behind  his (her) partner. Similar situations are represented in Fig.~\ref{crowd_scheme_a}. See, for example, the orange dyad and the gray individual.  

Besides, if $\varepsilon$ is strong enough to ensure that dyads will be at the contact distance, the scene will look like the one in Fig.~\ref{crowd_scheme_b}. No squeezing is now possible among the members of a dyad. These move together as a single bigger structure (the dyad) towards the target. The delays occur due to congestion (and friction) between dyads during the way out. Notice that this scenario corresponds to the “Closer-Is-Faster” (CIF) regime since higher $\varepsilon$ leads to more packed dyads, and therefore, is somehow more favorable to the evacuation process than the one in Fig.~\ref{crowd_scheme_a}.

Recall our observation in Section~\ref{section:results_fd} that strong attraction feelings reduce the “local effective density”. This becomes quite clear from Fig.~\ref{crowd_scheme_b}. The global density can stay moderate, say at 4-5$\,$P/m$^2$, although the social groups form closely packed structures. As a consequence, there is more space left between the groups, allowing a free-moving regime. 

The schemes represented in Fig.~\ref{crowd_scheme} resume the CIS and CIF phenomena. However, these may occur in different contexts. For instance, Fig.~\ref{dyad_interrumpted_a} shows the situation where a member of the purple dyad is trapped among the blocking cluster while the other is trying to cross the door.  Figs.~\ref{congested_corridor} and \ref{congested_corridor_gs5} are corresponding examples of weak and strong attractive feelings in a crowed corridor. 

We call the attention on the fact these phenomena can be obscured by other circumstances. We mentioned the groups' shape as one of these circumstances in Section~\ref{section:results_fd}. But the limiting walls (of the corridor or door sides) can be thought as another one (examine Figs.~\ref{dyad_interrumpted}, \ref{broken_links} and \ref{big_individuals} as examples). Further investigation will focus on these ones.

\section{Conclusions}\label{section:conclusions}

Our investigation focused on the evacuation of social groups in the context of the SFM. We upgraded the basic SFM according to the attractive feelings suggested in Ref.~\cite{FrankDorso}. We calibrated the model parameters with novel empirical results appearing in Ref.~\cite{dependence_dyad_dynamics}. We then focused on two scenarios: the bottleneck and the corridor situations. 

Our major result concerning the bottleneck scenario is that dyads (social groups of two members) slow down the emergency evacuations, although the final performance depends on the emotional involvement of the partners: moderately involved partners (\textit{colleagues}) yield the worst evacuation performance, while highly involved partners (\textit{intimate couples}) develop a better perfomance. We call both situations as the “Closer-Is-Slower” regime (for \textit{colleagues}) and the “Closer-Is-Faster” regime (for \textit{intimate couples}), respectively. 

The investigation on the nature of the delays in each regime showed that long delays are associated to some kind of disorder among the congested crowd. That is, moderately involved partners allow others to squeeze through them, and therefore, slow the evacuation down. \textit{Intimate couples}, instead, remain always together, keeping some kind of “order” during the evacuation.

We emphasize that the “Closer-Is-Slower” and “Closer-Is-Faster” regimes express some kind of parallelism with the known “Faster-Is-Slower” and “Faster-Is-Faster” regimes. We correlated all of them along the attractive intensities vs. anxiety levels in Section~\ref{section:delays}. 

The corridor scenario experiences two significant phenomena whenever social groups are present, which can be summarized as follows:  

(a) The free-flow regime extends towards more densely crowded situations than in the case of single individuals (to around 4-5$\,$P/m$^2$). This means that social groups yield to a more easily moving situation, as discussed in Section~\ref{section:discussion}.   

(b) In the highly congested regime, the flow slows down more pronouncedly than in the case of single individuals. This means that social groups do not move as smoothly as the single individuals in a highly dense environment. This was studied in Section~\ref{section:results_fd}. 

These phenomena become more noticeable when the groups consist of at least three members, and if these are highly emotionally involved.

In summary, our results show that socials groups tend to slow down the crowd dynamics (in the context of the SFM). However, the bottleneck and corridor situations also show that the final performance depends strongly on other considerations, say, the specific geometry where the situation takes place.

\section*{Acknowledgments}

This work was supported by the Agencia Nacional de Promoción Científica y Tecnológica, Argentina) grant FONCYT 2019 Number PICT-2019-2019-01994. G.A Frank thanks Universidad Tecnológica Nacional (UTN) for partial support through Grant PID Number SIUTNBA0006595.

\clearpage
\appendix
\setcounter{figure}{0}    
\section{Balance between social and attractive forces} \label{appendix:balance}

This appendix analyses the dynamic of an isolated dyad. It may be considered as an approach for low density crowds, where the dyad partners keep a respectful distance with others. 

\begin{figure}[bht]\centering
 \begin{subfigure}{.49\textwidth} \centering
 \includegraphics[width=\linewidth]{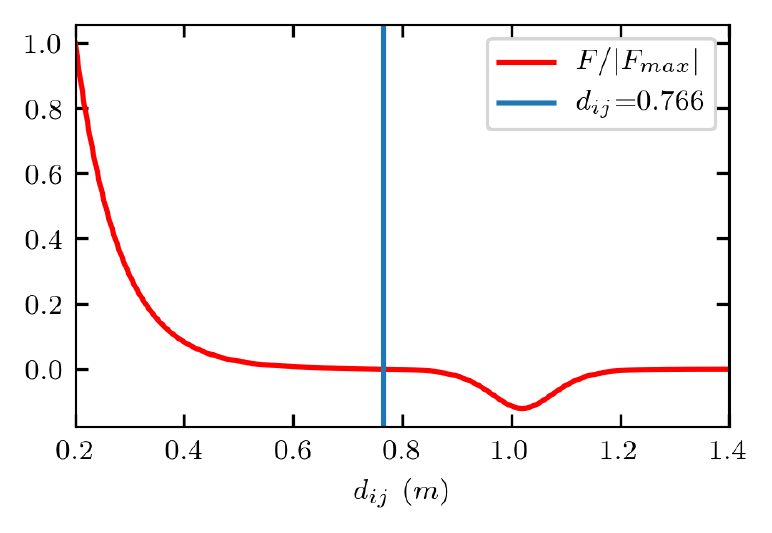}
 \caption{$\varepsilon=3$ }
 \label{atrac+social.a}
 \end{subfigure}%
 \begin{subfigure}{.49\textwidth} \centering
 \includegraphics[width=\linewidth]{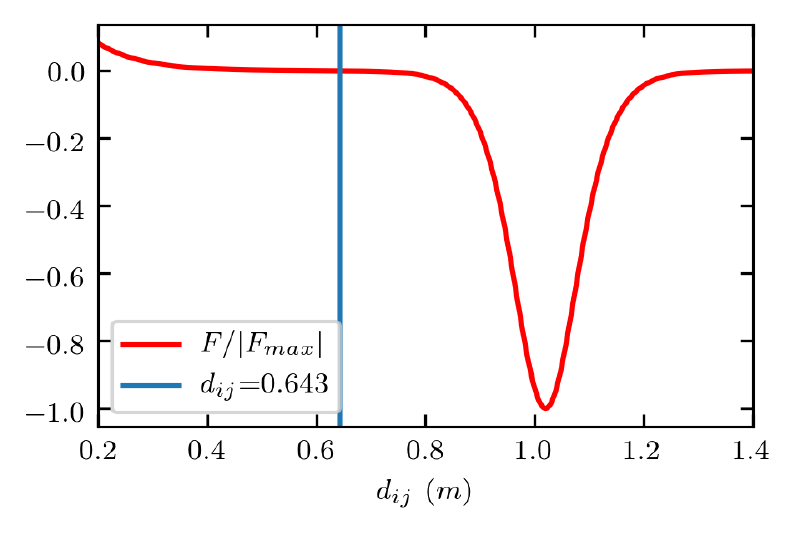}
 \caption{$\varepsilon=5$ }
  \label{atrac+social.b}
 \end{subfigure}
\caption{Net force between partners (sum of the social force and the attractive force). The blue line corresponds to $d_{ij}$ for which $|\mathbf f_s(d_{ij})|=|\mathbf f_a(d_{ij})|$.  \textbf{(a)} $\varepsilon=3$. \textbf{(b)} $\varepsilon=5$. }
\label{atrac+social}
\end{figure}

We will neglect the sliding friction between partners since we do not expect any relevant tangential velocity difference between them. Thus, partners will only interact via the social and attractive forces. Fig. \ref{atrac+social} shows the sum of these forces as a function of the distance between them, $d_{ij}$, for two values of $\varepsilon$. 

\FloatBarrier

We call $d_{eq}$ as  the equilibrium distance between partners. If $d_{ij}<d_{eq}$ the net force will be positive (repulsive force), and if $d_{ij}>d_{eq}$ it will be negative (attractive force). The balance equation may be approximated to the following first order expression
\begin{equation}
 f_s(d_{12})+f_a(d_{12}) \simeq -k(d_{12}-d_{eq})\qquad \text{ if } |d_{12}-d_{eq}|\ll 1   
\end{equation}
\noindent where $k$ depends on the parameters of the forces, like $A$ and $\varepsilon$. Notice that this is actually similar to an harmonic force equation.

In addition to the interaction forces, each individual is motivated to move to a target position through the “desired velocity” (see Section~\ref{section:SFM}). If the desired velocity is the same for both partners, say $\mathbf v_d=v_{d}\ \mathbf{\hat{e}}_d$, the equations of motion are as follows
\begin{equation*}
m\ddot{\mathbf {r}}_1=\mathbf f_a^{12}+\mathbf f_s^{12}+\mathbf f_d^1\simeq
-k\left( |\mathbf {r}_1-\mathbf {r}_2  |-d_{eq}   \right)\mathbf{\hat r}_{12}+m
\frac{ \mathbf v_d- \dot{\mathbf{r}} _1 }{\tau}
\end{equation*}
\begin{equation*}
m\ddot{\mathbf {r}}_2=\mathbf f_a^{21}+\mathbf f_s^{21}+\mathbf f_d^2\simeq
\textcolor{white}{-}k\left( |\mathbf {r}_1-\mathbf {r}_2  |-d_{eq}   \right)\mathbf{\hat r}_{12}+m
\frac{ \mathbf v_d- \dot{\mathbf{r}} _2 }{\tau}
\end{equation*}
Subtracting the bottom one to the top one, and defining $\mathbf{r \equiv r_1-r_2 }$, and $\omega^2\equiv2k/m$, leads to
\begin{equation}\label{ec_resorte}
\ddot{\mathbf{r}}=-\omega^2(|\mathbf{r}|-d_{eq})\mathbf{\hat {r}}-\frac 1\tau \ \dot{\mathbf{r}}\ 
\Rightarrow\ \boxed{\ 
\ddot{\mathbf{r}}+\frac 1\tau \ \dot{\mathbf{r}}+\omega^2 \mathbf{r} = \omega^2 d_{eq}\mathbf{\hat {r}\ }
}\end{equation}
This is the equation of a damped harmonic oscillator in two dimensions, with frequency $\omega$, decay time $2\tau$ and equilibrium distance $d_{eq}$. So, in a negligible amount of time compared to an egress time, each dyad reaches its equilibrium distance. It is interesting to note that this damping is caused by the desire force but does not depend on $v_d$, as the decay time is given by $\tau$. 

\begin{figure}[ht]\centering
 \includegraphics[width=0.6\linewidth]{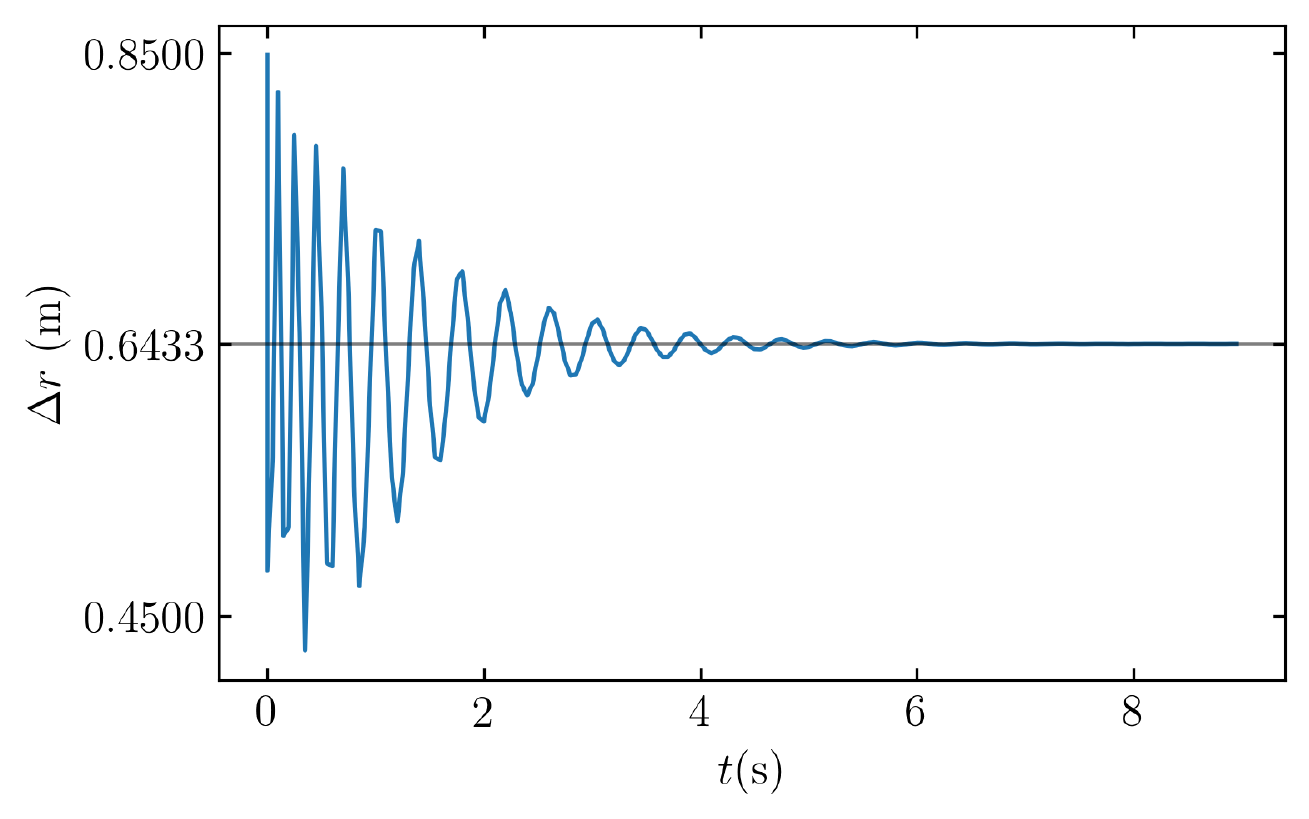}
 \caption{Distance between the members of a dyad in a simulation. The attractive intensity is $\varepsilon=5$. The initial separation is 0.84$\,$m.}
 \label{damped_oscilations}
\end{figure}

Fig. \ref{damped_oscilations} shows the distance between the members of a dyad as a function of time. Varying $A$, $\varepsilon$ and $v_d$, showed no significant differences in the decay time, and it only changed when varying $\tau$, as expected.

Notice that for $\varepsilon=5$, the equilibrium distance is the same as the one depicted in Fig. \ref{atrac+social.b}. Similarly, for any other $\varepsilon$, the equilibrium distance found from the balance equation $|\mathbf f_s(d_{eq})|=|\mathbf f_a(d_{eq})|$ and from low density simulations are the same. 
 
\FloatBarrier
\setcounter{figure}{0}    
\section{Evacuation time from low to moderate anxiety levels and very strong attractive feelings} \label{appendix:lowvdhigheps}

\begin{figure}[ht]\centering
 \begin{subfigure}{.325\textwidth} \centering
 \includegraphics[width=\linewidth]{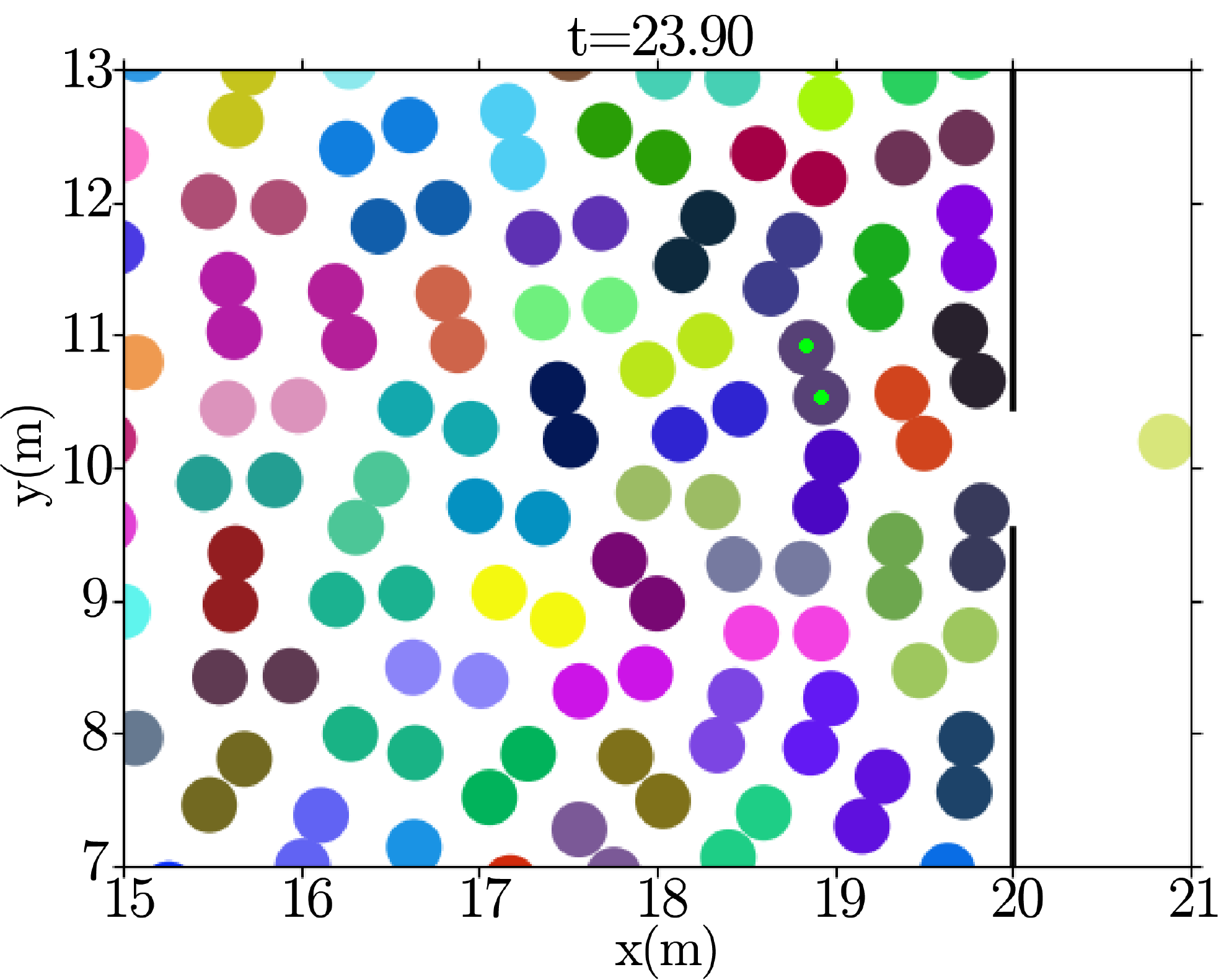}
  \end{subfigure}
  \begin{subfigure}{.325\textwidth} \centering
 \includegraphics[width=\linewidth]{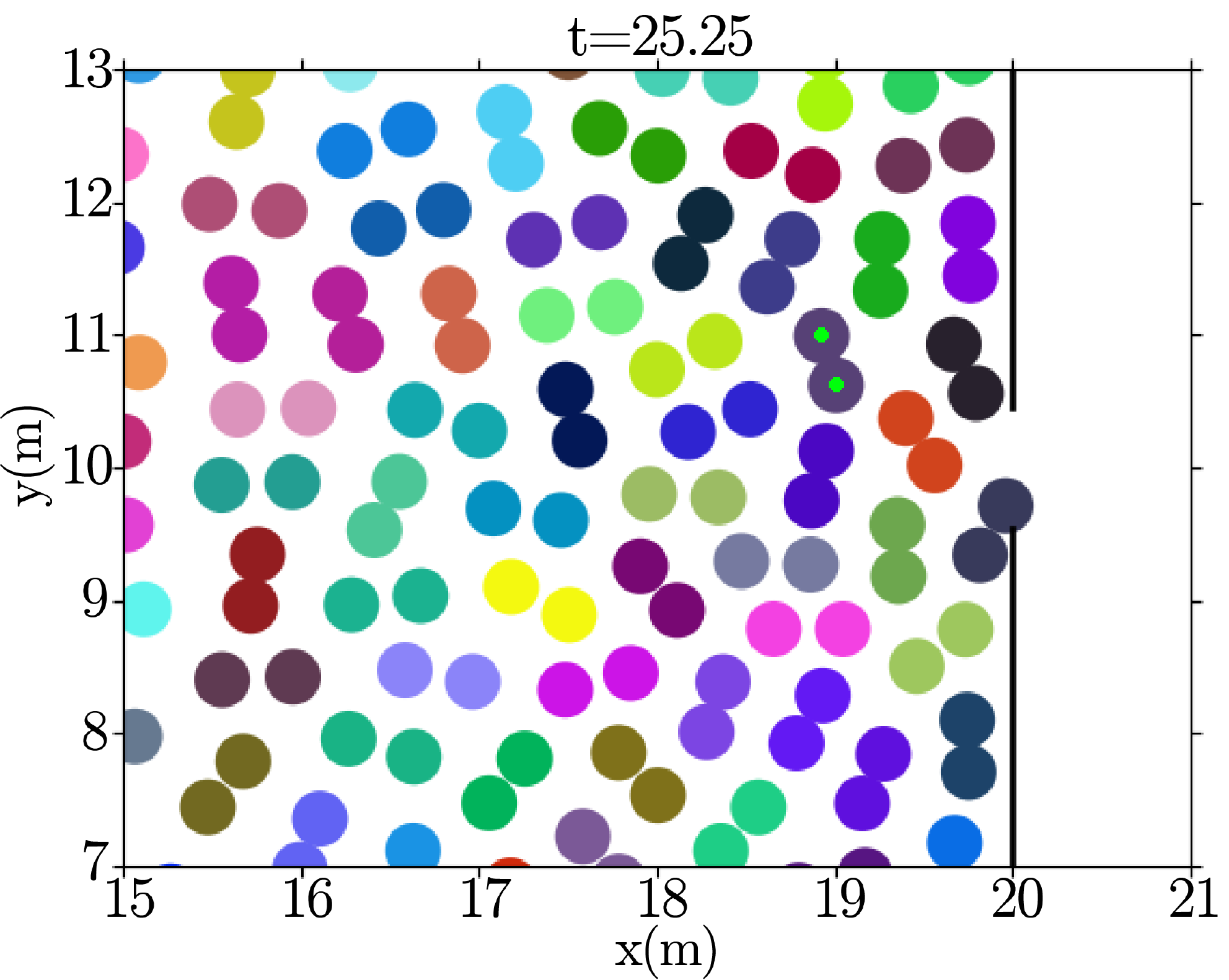}
  \end{subfigure}
  \begin{subfigure}{.325\textwidth} \centering
 \includegraphics[width=\linewidth]{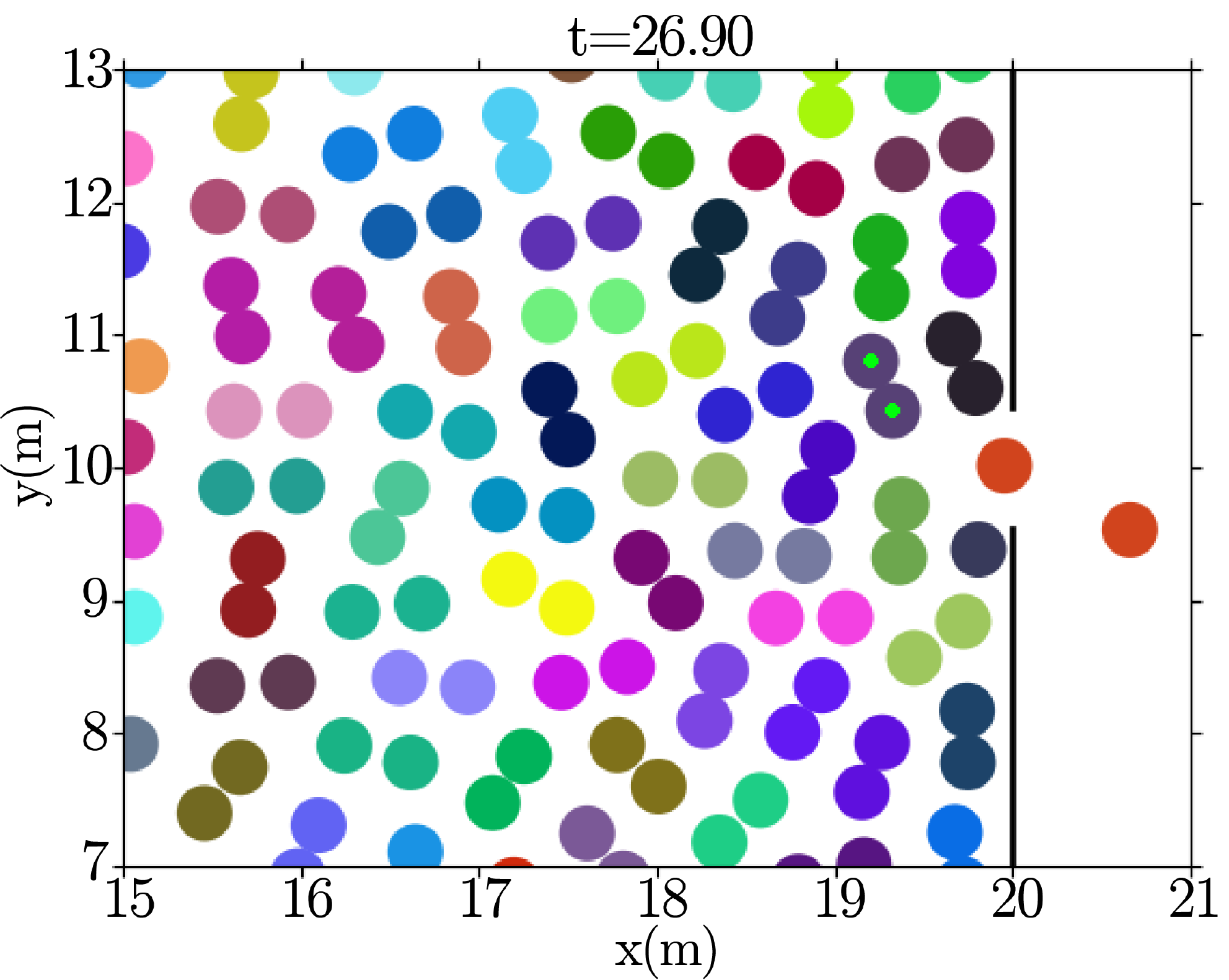}
  \end{subfigure}
  \begin{subfigure}{.325\textwidth} \centering
 \includegraphics[width=\linewidth]{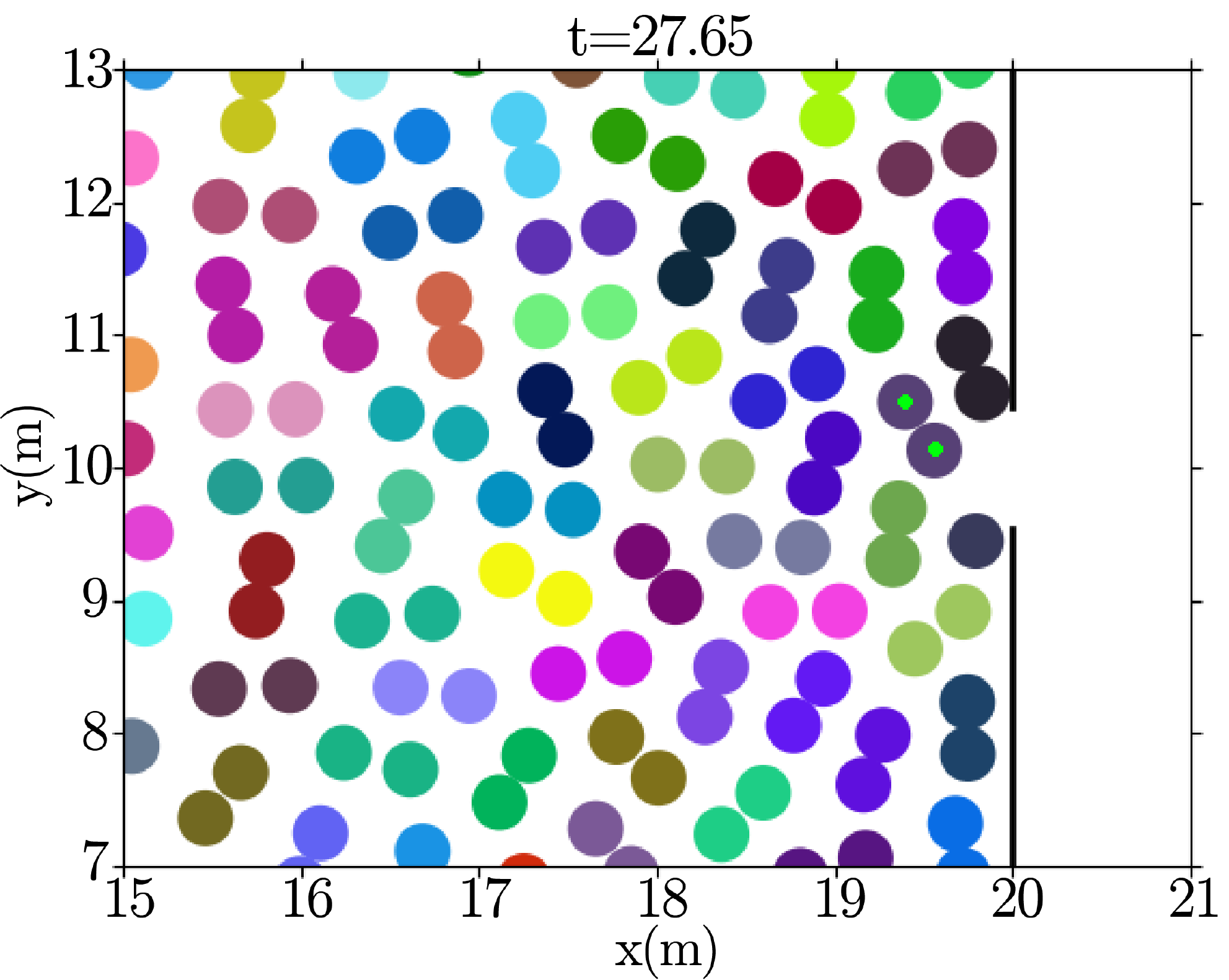}
  \end{subfigure}
  \begin{subfigure}{.325\textwidth} \centering
 \includegraphics[width=\linewidth]{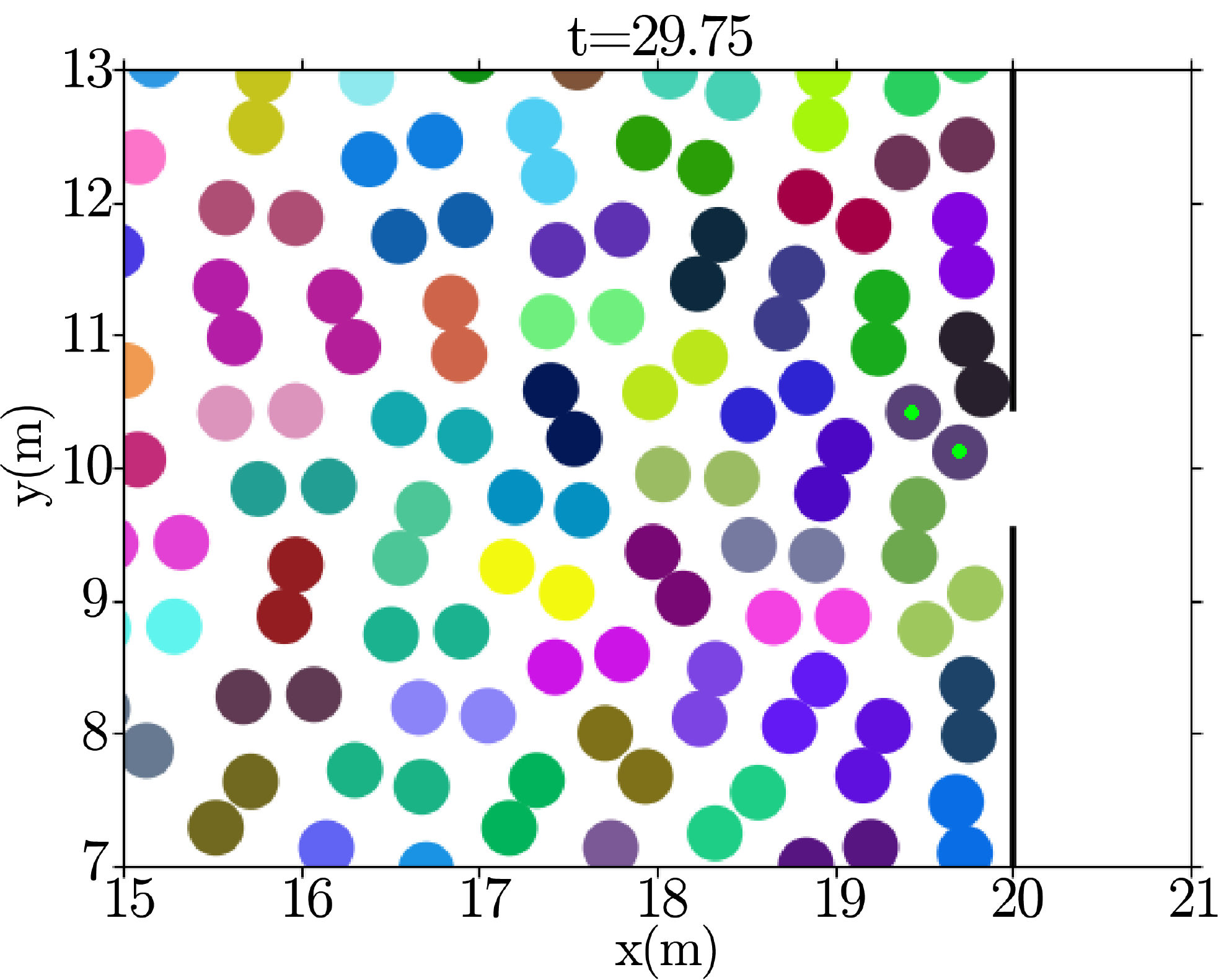}
  \end{subfigure}
  \begin{subfigure}{.325\textwidth} \centering
 \includegraphics[width=\linewidth]{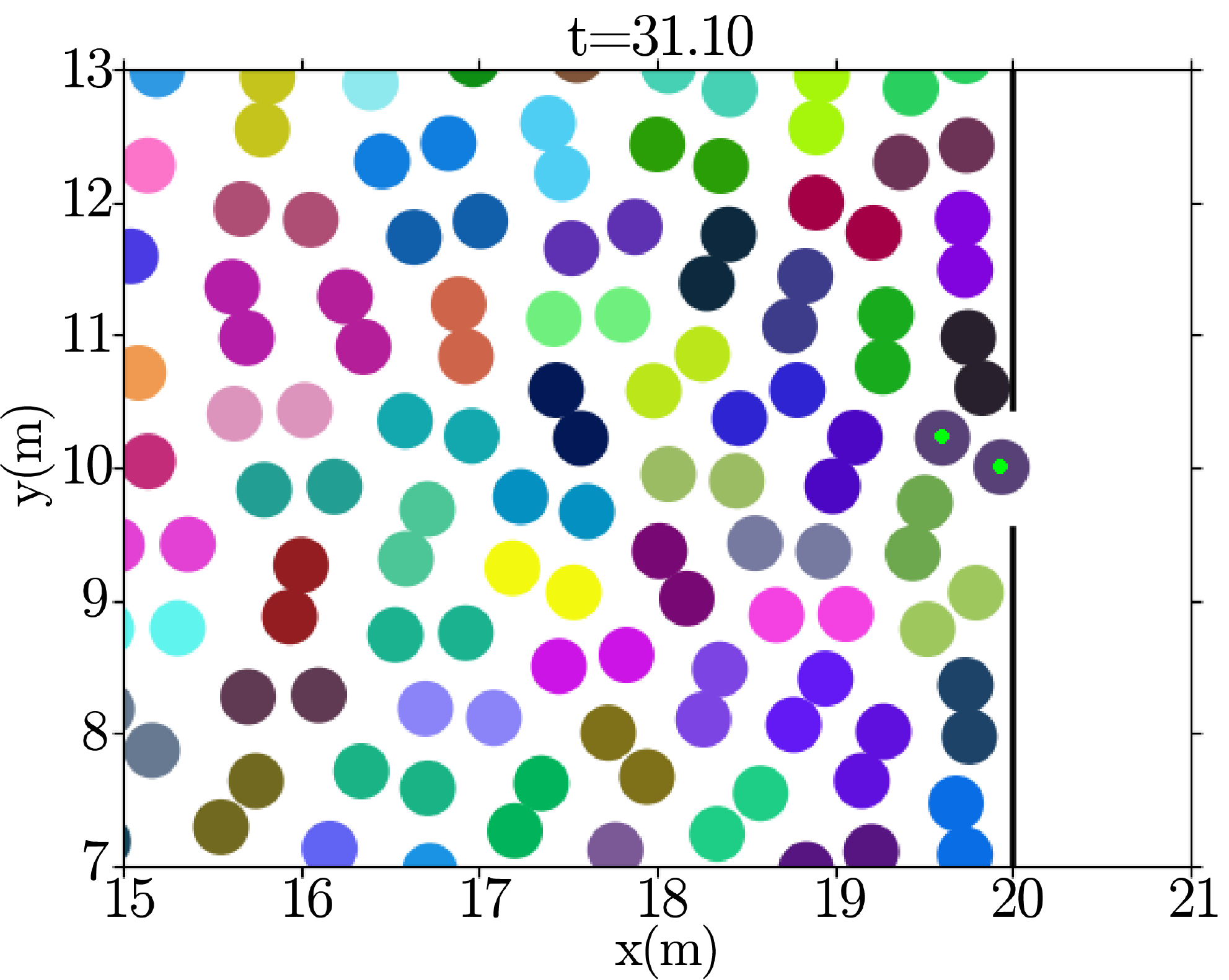}
  \end{subfigure}
 \caption{(Color online) Snapshots of an evacuation simulation process with $v_d=1.25\,$m/s and $\varepsilon=9$. Each dyad is represented by a different color. The \textit{purple dyad} with a \textit{green dot} is highlighted as an example of how dyads advance towards the exit (see text for details).}
\label{rigid_body}
\end{figure}

We notice in Figs.~\ref{t_heat_map} and ~\ref{delays_ind} (see Sections~\ref{section:t_evac} and \ref{section:delays}), that emergency evacuations with low anxiety levels ($v_d<1.5\,$m/s) slow down as $\varepsilon$ gets to $\varepsilon\simeq 8$. Video animations of the simulation processes show that dyads are permanently in contact during the evacuation. This means that partners behave as a single body. Fig.~\ref{rigid_body} captures an example of the animations.

For moderate and high $v_d$, we see no empty spaces surrounding the exit, as can be seen in Figs.~\ref{dyad_interrumpted} and \ref{broken_links}. Fig.~\ref{rigid_body} shows, however that the individuals prefer to stay with their partners instead of moving to the available spots near the exit. In other words, the attractive feelings surpass the desire to get out of the room, delaying both partners even when they have the option to advance.

The above behavior occurs because panic is not really present, and thus, other emotions dominate the scene. We do not focus on these situations since they are out of the emergency context.

\setcounter{figure}{0}    
\section{The velocity profiles of dyads in the corridor} \label{appendix:vel_profile}

\begin{figure}[ht]\centering
 \includegraphics[width=0.65\linewidth]{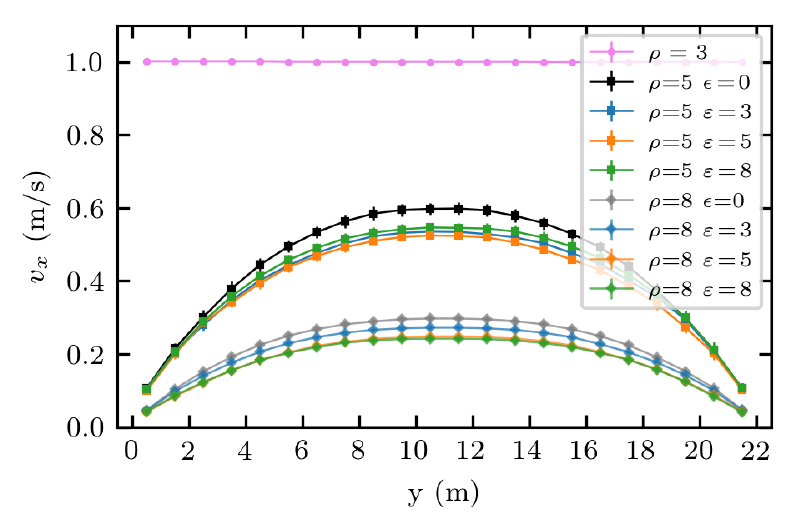}
 \caption{Velocity profile ($v_x$ vs. $y$) for various densities $\rho$ (shown in P/m$^2$ units) and $\varepsilon$. The simulated corridor is 28$\,$m long with periodic boundary condition along the walking direction, $x$. The corridor is delimited by walls at $y=0$ and $y=22\,$m.}
 \label{profile_dyads}
\end{figure}

To further examine the ordering and disordering in the corridor scenario, Fig.~\ref{profile_dyads} shows the velocity profiles (the velocity in the direction of the corridor as a function of the transverse coordinate, say, $v_x$ vs. $y$).

\FloatBarrier

For densities within the free-flow regime, all pedestrians move at the desired velocity ($v_x\!=\!v_d\!=\!1\,$m/s) for all $\varepsilon$, and thus the flow grows linearly as a function of the density (see Eq.~(\ref{flow_def})). As the density increases within the congested regime, the friction with the walls and between individuals slows the pedestrians down and generates a seemingly parabolic velocity profile. This result is similar to the velocity profiles for laminar flows in viscous fluids, where the velocity attains a maximum at the center and decreases towards the walls. This behavior has been studied before in Ref.~\cite{role_friction} and has also been observed in real crowds (see Ref.~\cite{unidirectional_dense_crowd}).

From Fig.~\ref{profile_dyads}, we can notice two different behaviors: 

\begin{itemize}
\item At $\rho=5\,$P/m$^2$, increasing $\varepsilon$ from 3 to 5 decreases the velocity, while increasing $\varepsilon$ to 8 increases it. This is consistent with the CIS and CIF regimes.
\item At $\rho=8\,$P/m$^2$, increasing $\varepsilon$ decreases the measured velocity, meaning that only the CIS regime is present. 
\end{itemize}

These behaviors can also be seen in the fundamental diagrams in Fig.~\ref{fd_dyads} at the mentioned densities. 

\begin{figure}[ht]\centering
 \includegraphics[width=0.7\linewidth]{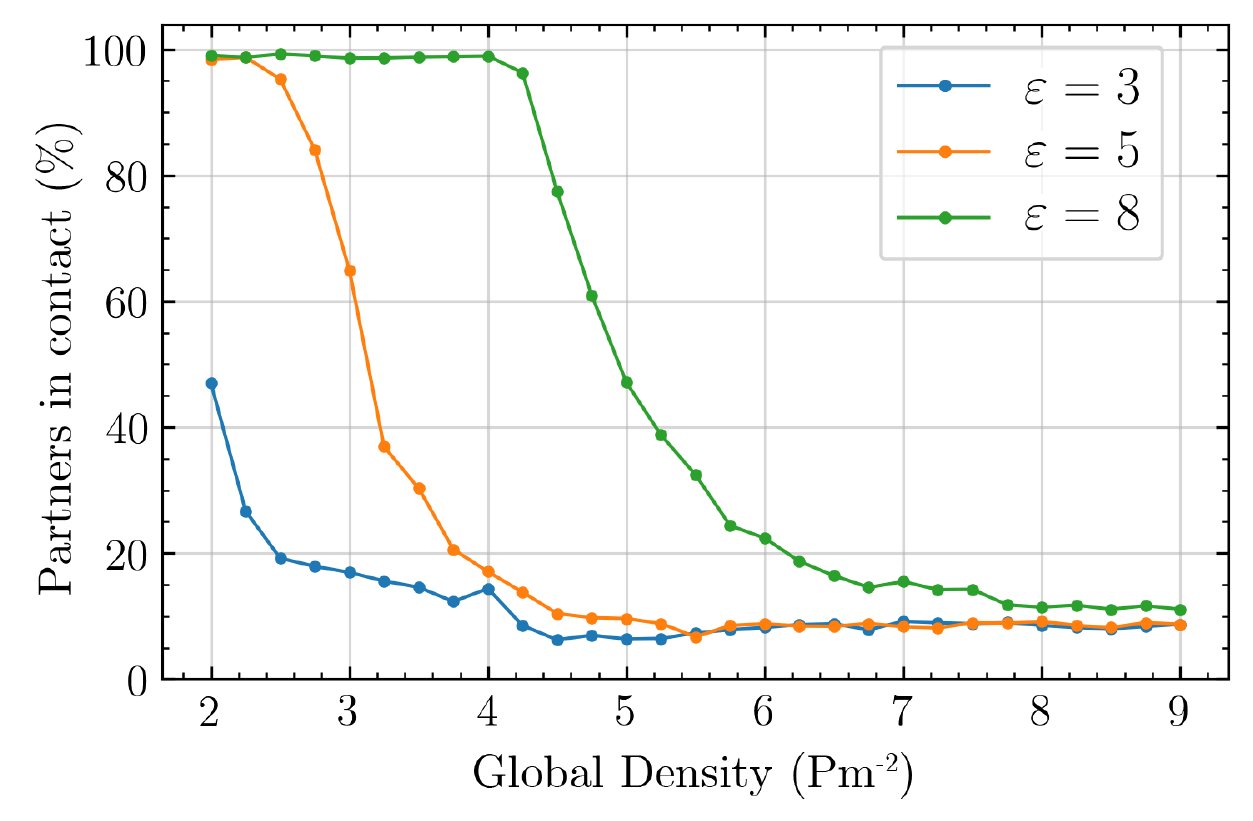}
 \caption{Fraction of dyads in contact as a function of the global density. The simulation conditions are the same as in Section~\ref{section:results_fd}, and measurements were taken after 50$\,$s of simulation.}
 \label{correlation}
\end{figure}

Fig.~\ref{correlation} shows the percentage of grouped pedestrians that have their dyad-partner as the closest individual to them. As the crowd gets more congested, it becomes harder for dyads to remain together. Notice that for higher $\varepsilon$ at a fixed density amount of partners at contact distance increases. For instance, only 10\% of the dyads are connected at $\rho=5\,$P/m$^2$ and $\varepsilon=3$ and 5. This scales to 50\% at $\varepsilon=8$. 

The above result supports our discussion in Section 6, say, that when dyad members are not physically in contact but are close enough to attract each other, the attractive forces will slow down the pedestrians' evacuation (see Fig.~\ref{crowd_scheme}). 

In summary, for each global density there is a range of $\varepsilon$ in which the dyads members tend to separate, and a different range in which the majority of them stay at contact distance. Within the first one, increasing $\varepsilon$ is not enough to significantly increase the amount of dyads that stay together, but it does increase the attractive forces that slow the flow down, thus this range coincides with the “Closer-Is-Slower” regime (see Fig.~\ref{crowd_scheme_a}). Similarly, in the second range increasing $\varepsilon$ increases the amount of dyads that stay together, reducing the slowing down of the flow (see Fig.~\ref{crowd_scheme_b}), therefore it corresponds to the “Closer-Is-Faster” range.

For very high densities, only the first regime is found. For example, for $\rho\geq 6.5\,$P/m$^2$ in Fig.~\ref{correlation}, even when $\varepsilon=8$ the percentage of connected dyads is below 20\%. Thus, only the “Closer-Is-Slower” regime is present. 

\begin{figure}[ht]\centering
 \includegraphics[width=0.85\linewidth]{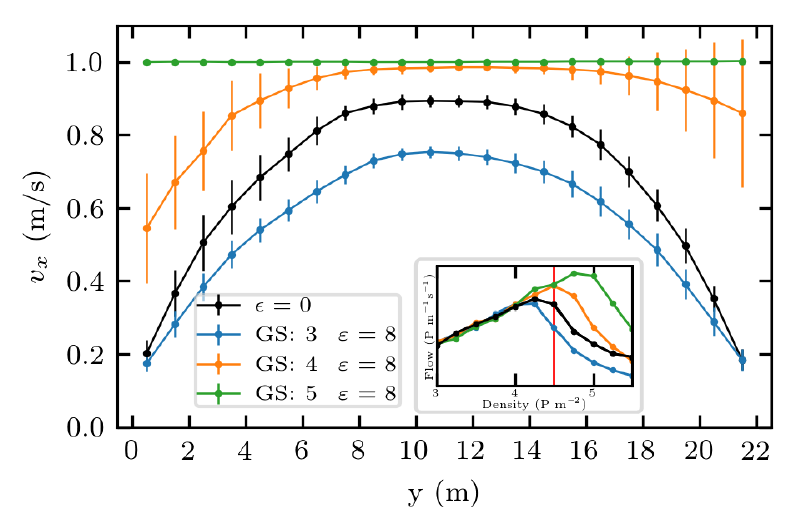}
 \caption{Velocity profiles for $\rho=4.5\,$P/m$^2$ in a corridor where 70\% of the pedestrians belong to groups of up to 5 individuals. The attractive intensity was $\varepsilon=8$. GS stands for the group size, and all groups were of the same size in each simulated process. The respective fundamental diagrams for densities near $4.5\,$P/m$^2$ are also shown for reference.  }
 \label{profile_intimate}
\end{figure}

The behavior is quite different for larger groups. Recall from Section~\ref{section:results_fd} that at $\rho$~=~4.5~P/m$^2$, the congestion degree varies depending on the group size. We further inspect the velocity profiles at this density value in Fig.~\ref{profile_intimate}  corresponding to each case. We can distinguish three patterns, as follows:

\begin{itemize}
    \item The fundamental diagram corresponding to crowds with no groups and with groups of 3 individuals is in the congested regime (black and blue curves, respectively), the velocity profile have a parabolic shape.
    \item With groups of 5 individuals (green curve), the free-flow regime is still present, and all the pedestrians move at their desired velocity $v_x~=~v_d~=~1$~m/s.
    \item With groups of 4 individuals (orange curve), the corridor is barely congested and the flow is transitioning between the two regimes. The velocity profile is thus in an intermediate stage, morphing from a straight line to a parable shape.
\end{itemize}

\FloatBarrier

\bibliography{refs}
\bibliographystyle{elsarticle-num}

\end{document}